\documentclass{aa}

\newcommand{\FDR}{\text{FDR}}
\newcommand{\FPR}{\text{FPR}}
\newcommand{\TPR}{\text{TPR}}
\newcommand{\FN}{\text{FN}}
\newcommand{\FP}{\text{FP}}
\newcommand{\TN}{\text{TN}}
\newcommand{\TP}{\text{TP}}
\newcommand{\zI}{z^{(i)}}

\newcommand{\yI}{y^{(i)}}
\newcommand{\xI}{x^{(i)}}
\newcommand{\best}[1]{#1}
\newcommand{\me}[2]{#1$^{\pm #2}$}
\usepackage{hyperref}
\usepackage[capitalize]{cleveref}
\crefname{figure}{Fig.}{Figs.}
\crefname{table}{Table}{Tables}
\crefname{equation}{Eq.}{Eqn.}
\Crefname{figure}{Figure}{Figures}
\Crefname{table}{Table}{Tables}
\Crefname{equation}{Equation}{Equations}
\crefname{section}{Sect.}{Sects.}
\Crefname{section}{Section}{Sections}

\usepackage{graphicx}
\usepackage{txfonts}

\begin{document} 

    \title{Evaluating the feasibility of interpretable machine learning for globular cluster detection}

   \subtitle{}
    
   \author{Dominik Dold\inst{*}\fnmsep\inst{1}
          \and
          Katja Fahrion\inst{*}\fnmsep\inst{2}
          }
    
   \institute{\inst{*} \textit{Both authors contributed equally to this work.}\\
   \inst{1}European Space Agency, European Space Research and Technology Centre, Advanced Concepts Team, Keplerlaan 1, 2201 AZ Noordwijk, the Netherlands.\\
   \inst{2}European Space Agency, European Space Research and Technology Centre, Keplerlaan 1, 2201 AZ Noordwijk, the Netherlands.\\
   \email{dominik.dold@esa.int, katja.fahrion@esa.int}
             }

   \date{\today}

  \abstract

\abstract{Extragalactic globular clusters (GCs) are important tracers of galaxy formation and evolution because their properties, luminosity functions, and radial distributions hold valuable information about the assembly history of their host galaxies. Obtaining GC catalogues from photometric data involves several steps which will likely become too time-consuming to perform on the large data volumes that are expected from upcoming wide-field imaging projects such as Euclid. In this work, we explore the feasibility of various machine learning methods to aid the search for GCs in extensive databases.
We use archival Hubble Space Telescope data in the F475W and F850LP bands of 141 early-type galaxies in the Fornax and Virgo galaxy clusters. Using existing GC catalogues to label the data, we obtain an extensive data set of 84929 sources containing 18556 GCs and we train several machine learning methods both on image and tabular data containing physically relevant features extracted from the images. 
We find that our evaluated machine learning models are capable of producing catalogues of similar quality as the existing ones which were constructed from mixture modelling and structural fitting. 
The best performing methods, ensemble-based models like random forests and convolutional neural networks, recover $\sim 90-94$ \% of GCs while producing an acceptable amount of false detections ($\sim 6-8$\%) -- with some falsely detected sources being identifiable as GCs that have not been labelled as such in the used catalogues. In the magnitude range 22 $<$ m4\_g $\leq$ 24.5 mag, 98 - 99\% of GCs are recovered.
We even find such high performance levels when training on Virgo and evaluating on Fornax data (and vice versa), illustrating that the models are transferable to environments with different conditions such as different distances than in the used training data.
Apart from performance metrics, we demonstrate how interpretable methods can be utilised to better understand model predictions, recovering that magnitudes, colours, and sizes are important properties for identifying GCs. Moreover, comparing colour distributions from our detected sources to the reference distributions from input catalogues finds great agreement and the mean colour is recovered even for systems with less than 20 GCs.
These are encouraging results, indicating that similar methods trained on an informative subsample can be applied for creating GC catalogues for a large number of galaxies, with tools being available for increasing the transparency and reliability of said methods.}

   \keywords{galaxies: star clusters: general -- methods: data analysis -- galaxies: formation -- galaxies: evolution}

   \maketitle
%

\section{Introduction}

Globular clusters (GCs) are massive, dense star clusters that can be found in almost all galaxies (see the reviews by \citealt{BrodieStrader2006}, \citealt{Forbes2018a}, and \citealt{Beasley2020}). With typical masses between $10^4$ and $10^6 M_\sun$, and compact sizes (effective radii of $\sim 3 - 10$ pc), GCs are extremely dense stellar systems that are detectable in distant galaxies.

In the last decades, photometric surveys have produced extensive catalogues of GC candidates (e.g. \citealt{Jordan2007, Jordan2015, Cantiello2020}). These catalogues typically report coordinates, luminosities, and colours in at least two bands for individual GCs and enable exploration of a large variety of science cases related to the host galaxies and their assembly. 
For example, the luminosity or mass function of GC systems have long been used as distance indicators (e.g. \citealt{Richtler2003, LomeliNunez2022}).
Additionally, it has been established that the number of GCs is tightly correlated with the total dark matter halo mass of a galaxy \citep{Harris2017b, Forbes2018}, which is often interpreted as a consequence of hierarchical merging (e.g. \citealt{Valenzuela2021}). Similarly, the radial profiles of GC systems correlate with dark matter halo properties (e.g. \citealt{Hudson2018, ReinaCampos2021, DeBortoli2022}.
Moreover, GC colours are regarded as valuable indicators of a galaxy's merger history. Many galaxies exhibit a bimodal colour distribution with a blue and red population (e.g. \citealt{BrodieStrader2006}) that are often interpreted as stemming from two populations with different metallicities of different origins: a metal-poor, blue population that was born in now accreted dwarf galaxies and a red, metal-rich population formed in-situ in the host galaxy (e.g. \citealt{AshmanZepf1992, Cote1998, Beasley2002}). Although this strict division of accreted and in-situ formed GCs might a bit too simplistic (e.g. \citealt{ForbesRemus2018, Fahrion2020c}), GC colours are nonetheless valuable tracers of the underlying host metallicity (e.g. \citealt{GeislerLeeKim1996, ForbesForte2001, Fahrion2020b}). 

Due to their wide applicability, obtaining GC catalogues remains a relevant task for ongoing studies covering an ever-growing number of galaxies. However, the techniques to obtain these catalogues have remained the same for many years and usually involve multiple steps, including source detection, cleaning based on photometric properties, and magnitude and colour cuts (e.g. \citealt{Jordan2007, Harris2017, Cantiello2020, LomeliNunez2022}).
Works based on ground-based photometry usually rely on a large number of photometric filters as most extragalatic GCs appear unresolved in seeing-limited data, requiring that the colour-colour space is explored to remove contaminants such as background galaxies and foreground Milky-Way stars (e.g. \citealt{DAbrusco2016, Cantiello2018}). In contrast, photometric surveys with the \textit{Hubble} Space Telescope (HST) have exploited the fact that GCs are marginally resolved in HST data out to many tens of megaparsecs and thus require less photometric bands \citep{Jordan2007, Jordan2015}. However, in these studies the individual sources need to be modelled to infer their sizes which can be time consuming for a large number of initial detections. 

For upcoming space-based wide-field survey facilities such as Euclid \citep{Euclid} or the Nancy Grace Roman Space Telescope \citep{NancyGraceRoman}, such classical techniques might not be the best choice due to the amount of data that has to be analysed. For this reason, it is necessary to devise and test new methods of obtaining GC catalogues. In this paper, we apply and test a range of machine learning  techniques that are capable of handling large data sets on galaxies with existing photometric GC catalogues to explore the performance with regard to classical methods. 

In the recent years, many authors have approached astrophysical problems with machine learning, showing the general interest in such techniques for today's and future applications. To just name a few recent examples, machine learning has been used for classification of supernovae (e.g. \citealt{Fremling2021}), inference of galaxy halo properties \citep{vonMarttens2021, VillanuevaDomingo2021}, cosmological predictions \citep{Li2021, VillaescusaNavarro2021}, or galaxy identification and classification (e.g. \citealt{Mueller2021, Tarsitano2021, Ciprijanovic2021}).

Additionally, machine learning has been applied for star cluster science, both for individual galaxies and with data of larger surveys.
\cite{Bialopetravicius2019} tested convolutional neural networks (CNNs) on mock images of artificial resolved clusters and \cite{Wang2021} applied CNNs to identify star cluster candidates in M\,31. Additionally, \cite{Bialopetravicius2020a, Bialopetravicius2020b} explored CNNs to identify and study young star clusters in M\,83 from multi-band HST observations.
\cite{Wei2020} and \cite{Whitmore2021} demonstrated that deep learning can be used to identify 
young star clusters in HST data of the Physics at High Angular Resolution in Nearby GalaxieS (PHANGS)-HST survey, while \cite{Perez2021} presented a machine learning pipeline based on CNNs to identify (young) star clusters in HST data from the Legacy Extra Galactic  Ultraviolet  Survey (LEGUS). These studies apply various machine learning methods with success and typically report \mbox{$\gtrsim$ 90\%} success fractions for identifying bright, symmetric star clusters.
While the aforementioned works focused on young or resolved star clusters, identification of (old) GCs with machine learning methods has been recently studied by \cite{Mohammadi2022}. In this work, random forest (RF) and Localized Generalized Matrix Learning Vector Quantization (LGMLVQ) classifiers are used to identify $\sim$ 500 GCs and ultra-compact dwarf galaxies in multi-wavelength ground-based photometric data of $\sim$ 7700 sources in the Fornax galaxy cluster with promising results. 

In this paper, we aim to extend previous work on star cluster identification by applying and testing the performance of a range of different machine learning methods on space-based HST data of non-star forming galaxies for which GC catalogues exist. We assemble a large data set of $\sim$ 85000 sources containing $\sim$ 18500 known GCs, which was obtained by combining data from two different environments, the Virgo and the Fornax galaxy clusters.
The machine learning methods are thoroughly investigated on different data representations, such as image data and tabular features extracted from images, and we test different scenarios, for example training and testing on sources of different galaxy clusters.
We further explore methods from the field of explainable artificial intelligence, identifying that our models use similar indicators for detecting GCs as commonly employed in traditional methods.
The obtained results demonstrate that modern machine learning techniques are an intriguing tool for generating GC catalogues from upcoming space-based wide-field surveys.

In the following section, the data are explained in more detail. Section \ref{sect:methods} gives an overview of the used machine learning methods for tabular and image data. The results are presented in Sect. \ref{sect:results} and discussed in Sect. \ref{sect:discussion}. We conclude in Sect. \ref{sect:conclusions}.

\begin{figure}
    \centering
    \includegraphics[width=0.45\textwidth]{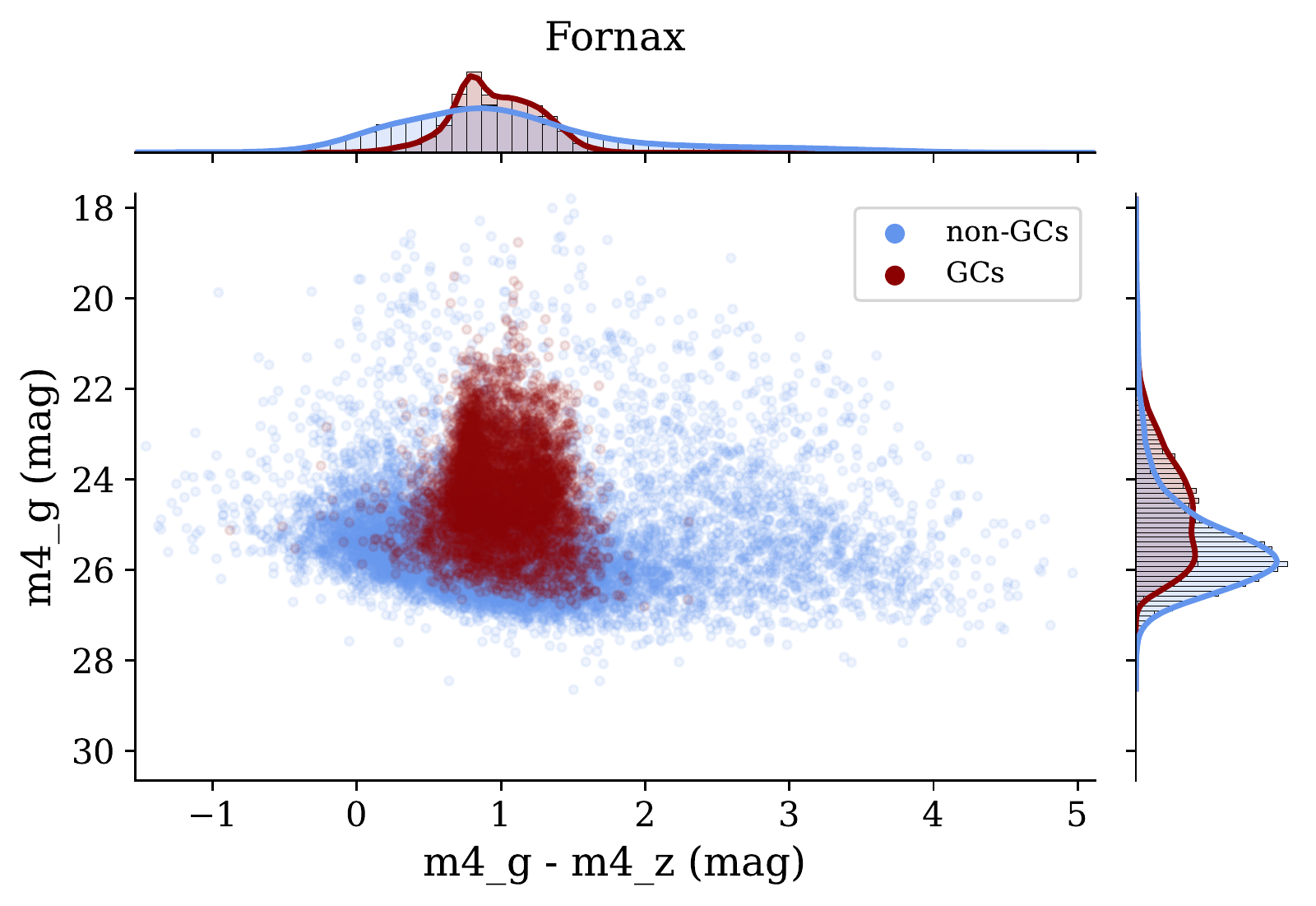}
    \includegraphics[width=0.45\textwidth]{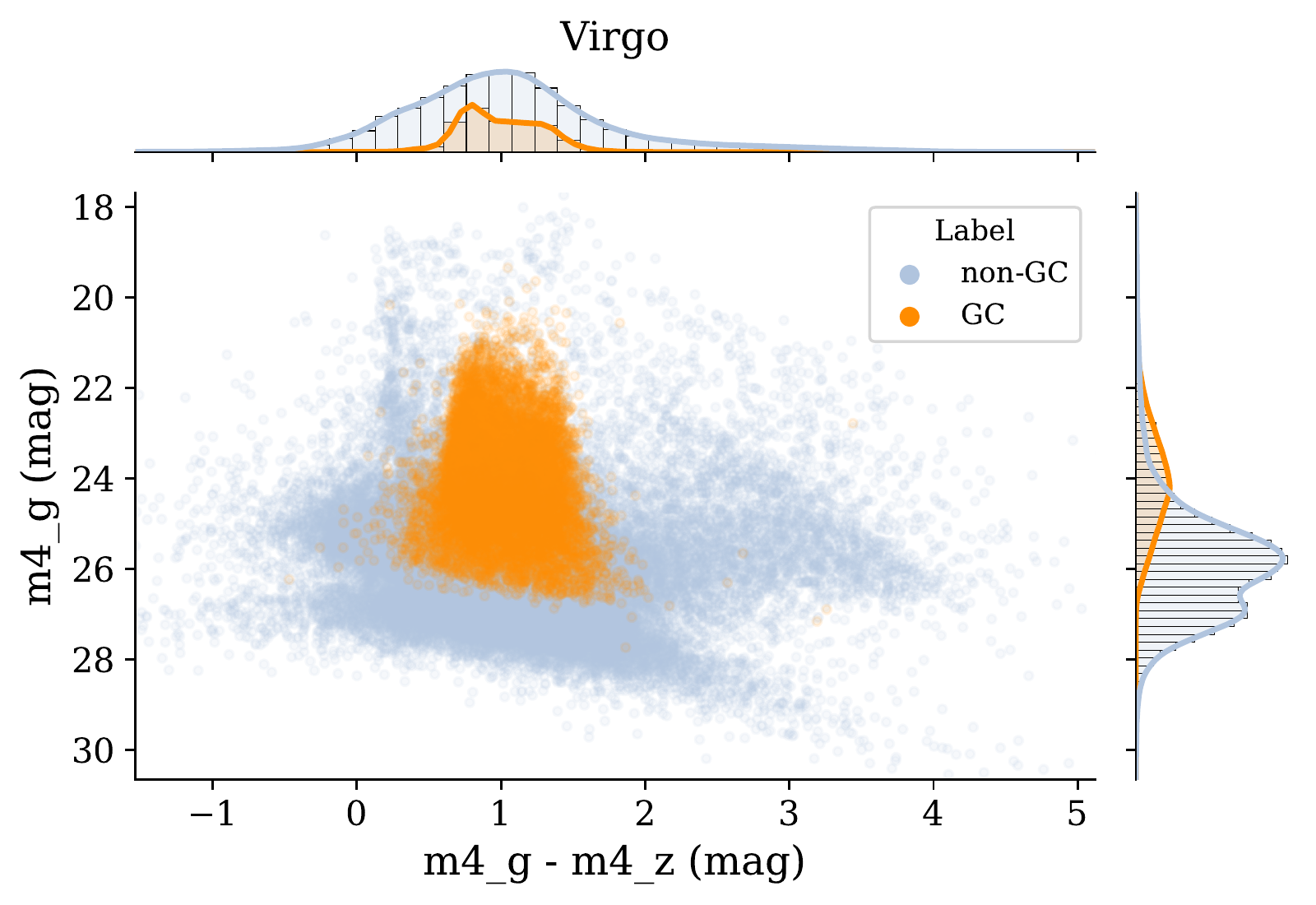}
    \includegraphics[width=0.45\textwidth]{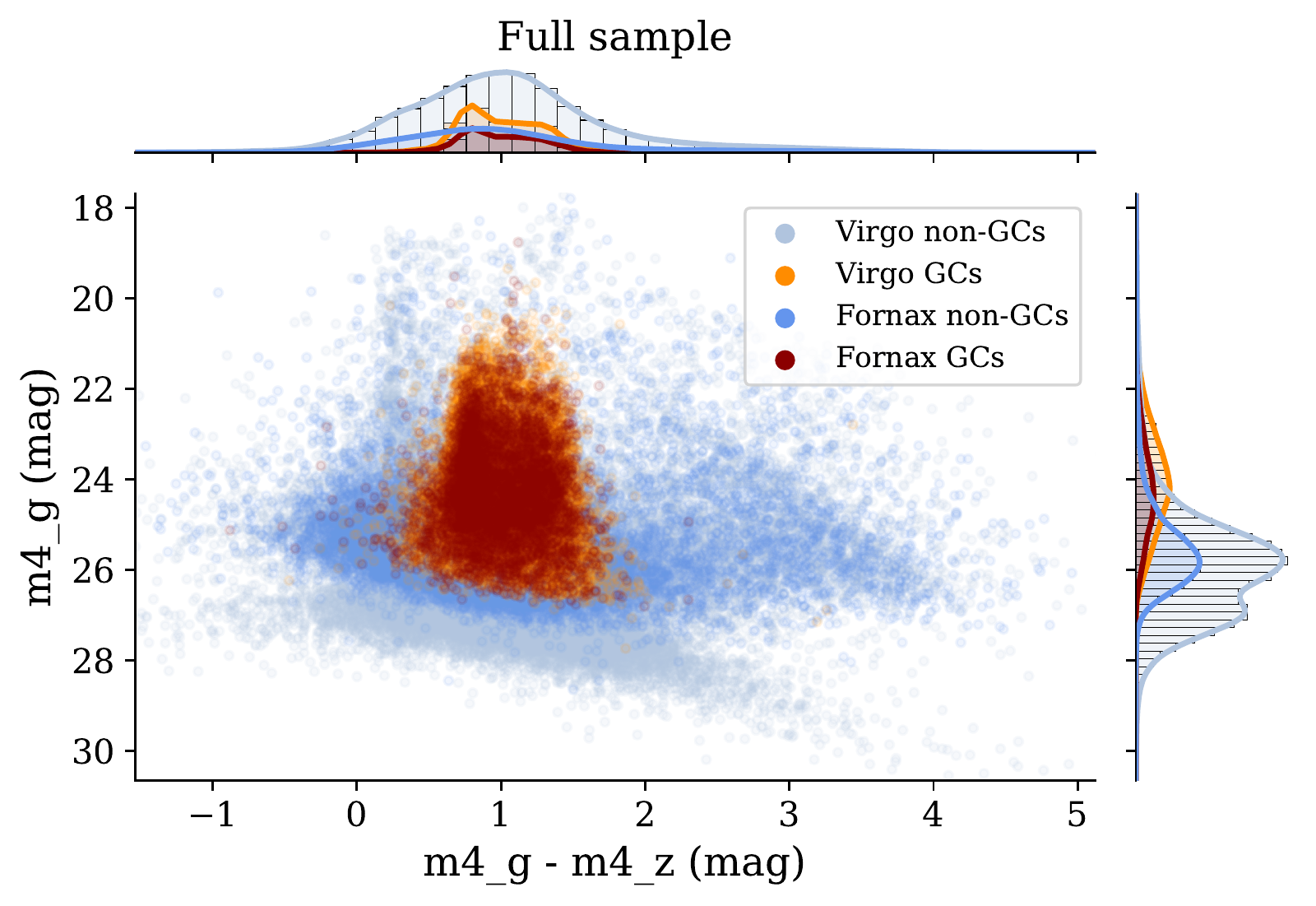}
    \caption{Colour-magnitude diagrams for the different samples based on 4 pixel aperture magnitudes that are corrected for foreground extinction. We differentiate between GCs as the ones that were matched to the ACSVCS and ACSFCS catalogues and non-GCs.}
    \label{fig:sample}
\end{figure}

\section{Data}
\label{sect:data}
We use the data from the Advanced Camera for Surveys (ACS) Virgo Cluster Survey (ACSVCS; \citealt{Cote2004}) and ACS Fornax Cluster Survey (ACSFCS; \citealt{Jordan2007a}), in which GCs are marginally resolved due to the close distances of the Virgo and Fornax galaxy clusters (16.5 Mpc and 20 Mpc, respectively; \citealt{Mei2007, Blakeslee2009}). GC catalogues were presented in \cite{Jordan2007} and \cite{Jordan2015}.  
Both the ACSVCS and the ACSFCS are surveys based on HST ACS observations in the F475W ($\sim g$) and F850LP ($\sim z$) bands of 100 massive early-type galaxies in Virgo and 43 galaxies in the Fornax galaxy cluster, respectively. We downloaded all available ACS data from the Hubble Legacy Archive\footnote{\url{https://hla.stsci.edu}} consisting of 98 galaxies in Virgo and all 43 galaxies in Fornax. For VCC\,1049 and VCC\,1261 no data were available. In the ACSVCS, the data have exposure times of 750 and 1210 seconds in the $g$- and $z$ bandpasses, respectively, and 760 ($g$-band) and 1220 ($z$-band) seconds in the ACSFCS.

\subsection{Existing globular cluster catalogues}
Extensive GC catalogues based on the ACS data exist for both galaxy clusters, presented in \cite{Jordan2009} for Virgo and \cite{Jordan2015} for Fornax. These catalogues were constructed using the methods detailed in \cite{Jordan2009} and are based on data reduction procedures and initial selections for the ACSVCS presented in \cite{Jordan2004a} and \cite{Jordan2007a} for ACSFCS.

Here, we summarise the most important steps. As described in \cite{Jordan2004a}, initial source catalogues were derived with SExtractor \citep{Bertin1996} after a surface brightness model of each galaxy was subtracted from the images. Then, initial cuts on the source magnitudes and shape were made, which excluded overly bright and elongated sources. Additionally, sources found within 0.5\arcsec of the galaxy centres were omitted to not include nuclear star clusters. Then, all sources were fitted with point-spread function-convolved King models \citep{King1966} to obtain structural and photometric parameters such as magnitudes, half-light radii, and the concentration $c$.

This initial sample still contained contaminants from faint foreground stars and background galaxies which were further removed based on a broad colour cut of $0.5 < (g - z) < 1.9$ mag and removal of extended as well as unresolved sources with 0.75 pc $< r_h <$ 10 pc. Then,  model-based clustering was used to divide the remaining sources into two components. While the contaminant component is assumed to be a fixed component based on a control field, the GC component model considers source magnitudes and sizes. Based on these parameters, the probability that a source is a GC $p_\text{GC}$ can be calculated. While the ACSVCS GC catalogue presented in \citep{Jordan2009} only contains sources with $p_\text{GC} \geq 0.5$, a slightly different selection was made for the ACSFCS GC catalogue \citep{Jordan2015} which also contains sources with $p_\text{GC} < 0.5$. 

In total, the ACSVCS catalogue contains 12763 GC candidates, the ACSFCS catalogue contains 9136 sources, 6275 of them with $p_\text{GC} \geq 0.5$. In the following, we will only use those as GC candidates. In summary, this thorough work of detecting, cleaning, and modelling the sources in the ACSVCS and ACSFCS projects has produced a extensive data set of 19038 GCs.

\subsection{Data preparation}
While the many steps that were taken to obtain the ACSVCS and ACSFCS GC catalogues have produced a rather clean set of GCs, in this work we wish to explore how well different machine learning methods can reproduce these GC catalogues starting from an unprocessed initial data set that was produced by simple source detection methods. For this reason, we derived our own source catalogues from the archival ACS data and only use the GC catalogues to label some of the detected sources as GCs. The data that are fed to the machine learning methods have not gone through any colour or magnitude cuts and also contain contaminants of many kinds, such as foreground stars, background galaxies, and image artefacts. 

To obtain our data, we take the following steps for each of the ACSFCS and ACSVCS galaxies.
First, a median filter is used to remove the smooth galaxy background using a 7 $\times$ 7 pixel filter in both bandpasses. Testing with different filter sizes gives similar results. Then, sources were detected in the residual images using Photutil's segmentation routines \cite{photutils}. We used a 3 $\times$ 3 pixel Gaussian kernel for filtering and selected all sources that have at least 2 pixels which are 3 $\sigma$ above the local median background. Standard deblending parameters with 32 levels were chosen. From this initial source list, we only keep those that were detected in both filters and that are at least 5 pixels from the edges of the detector gap that bisects the two ACS charge-coupled devices (CCDs) of the wide field channel (WFC).

Then this source list is matched with the ACSFCS and ACSVCS catalogues, labelling matched sources as GCs. In a last step, simple photometry is performed to get a measurement of the aperture magnitudes. We measured the magnitudes for each source using 3, 4, and 5 pixel apertures and corrected those for foreground extinction using the dust maps from \cite{Schlegel1998}. Additionally, we measured the concentration index CI of these apertures as the magnitude difference between each 3, 4, or 5 pixel aperture and a 1 pixel aperture. In the following, we use only these aperture magnitudes and do not apply any aperture correction as those would need additional information on the type of object that is observed. However, such corrections will need to be made eventually if the data is finally used for GC science (e.g. \citealt{Jordan2015}). The extracted features are described in Appendix \ref{app:features}.

\begin{figure}
    \centering
    \includegraphics[width=0.45\textwidth]{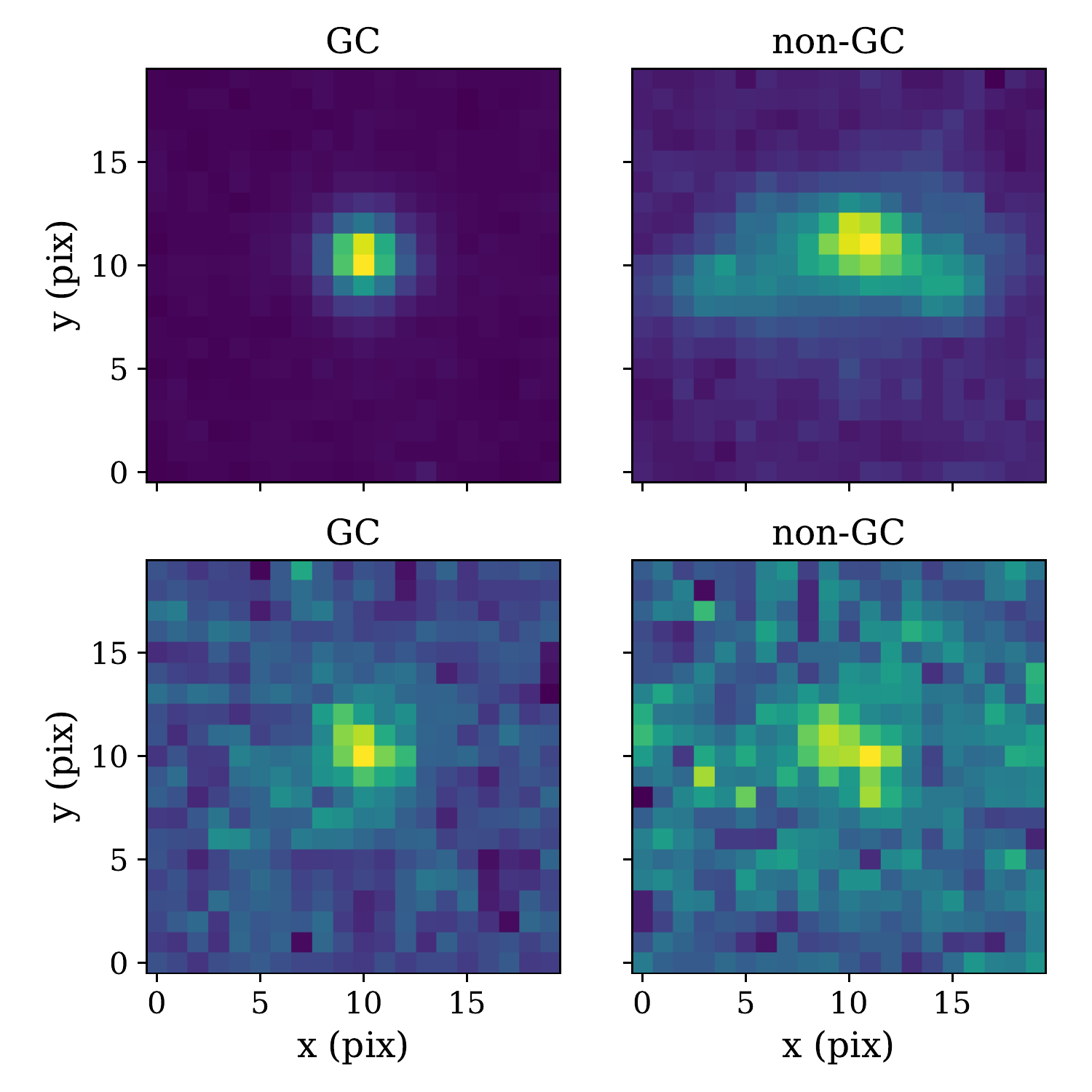}
    \caption{Examples of image data of GCs (left) and non-GCs (right). The sources have different magnitudes and the plotted flux scale differs from source to source.}
    \label{fig:example_globs}
\end{figure}

Unfortunately, not all sources from the ACSVCS and ACSFCS catalogues were detected in our simple pipeline. In Virgo, 304 (2.4 \%) of 12699 catalogue GCs\footnote{This number is lower than the one quoted above because we could not consider VCC\,1261 and VCC\,1049.} could not be found, 69 of those fainter than $g = 26$ mag. Often, those faint sources were only found in one filter.  114 GCs of 6275 (1.8 \%) were not found in the Fornax galaxies. 

In total, our source lists contain 84929 sources, 63162 in Virgo and 21767 in Fornax. 
Out of those, 18556 are matched 
to GC candidates in the ACSFCS and ACSVCS catalogues, 12395 in Virgo and 6161 in Fornax. Consequently, 21.8 \% of our sources are matched to catalogue sources and are labelled as GCs. The remaining sources are labelled as non-GCs. \Cref{fig:sample} illustrates this sample in colour-magnitude diagrams. The data set statistics are summarised in \cref{tab:datasets}.

As visible from Fig. \ref{fig:sample}, the colour-magnitude distributions for the sources matched with catalogue GCs are very similar between Virgo and Fornax, but the contaminant distributions vary slightly. One reason for this is the larger distance to Fornax (20 Mpc instead of 16.5 Mpc). In combination with the almost equal exposure time, this creates a cut-off of detected sources at brighter magnitudes than in Virgo. Additionally, the non-GC sources in Virgo show a small overdensity at m4\_g $-$ m4\_z $\sim$ 0.2 mag. Inspecting other parameters and the image cut-outs of these sources finds them to be dominated by compact sources with low concentration indices (CI $\sim$ 1). We therefore believe that these sources are foreground stars, which are not visible in the Fornax sample because of their overall low numbers ($< 0.5 \%$ of Virgo sources).

In addition to tabular feature data, we want to test machine learning methods on image data of the sources. For this reason, we created 20 $\times$ 20 pixel cut-outs in the $g$- and $z$-bands of all detected sources from the residual images which were previously used for source detection. The source centroid was placed into the centre of each image.
\Cref{fig:example_globs} shows four examples of such cut-outs. This illustrates that while some elongated sources (like in the top right panel) can be excluded as GCs easily by eye, the case becomes more difficult for fainter sources.
\begin{table}[hb]\renewcommand{\arraystretch}{1.2}
\caption{Statistics of the assembled data sets.}
\label{tab:datasets}
\center
\begin{tabular}{ c |  c  c  c  c}
 Data set & \# Sources & \# GCs & \# non-GCs & Class ratio\\
 \hline \hline
Fornax & 21767 & 6161 & 15606 & $\sim$1:2.5 \\
Virgo & 63162 & 12395 & 50767 & $\sim$1:4 \\
\hline
Total & 84929 & 18556 & 63829 & $\sim$1:3.5 \\
\hline
\end{tabular}
\end{table}

\section{Methods}
\label{sect:methods}
To get a thorough understanding of our data set and the problem of extracting GCs from photometric surveys, we evaluate the capability of several machine learning models with different degrees of complexity.
We apply such models on tabular features extracted from the images that are commonly used to identify GCs as well as the images directly. 
In the following, we briefly review the used methods as well as the evaluation scenarios and metrics used throughout the remainder of this document. 
A detailed mathematical description of the considered models can be found in \cref{app:models}.

\subsection{Notation}

We employ a uniform notation to describe both the data and the models.
By $\xI \in \mathbb{R}^N$, we denote the tabular feature vector of the $i$th source in our training data set, while $\yI \in \{0,1\}$ is the corresponding label with $\yI = 1$ if source $i$ is a GC and $\yI = 0$ otherwise.
$x \in \mathbb{R}^N$ denotes feature vectors in general.

Machine learning models are described as functions $p: \mathbb{R}^{N} \to \mathbb{R}$ that map from the feature space into the real space, from which a class label $y_\mathrm{m}(x)$ is inferred via thresholding
\begin{equation}
    y_\mathrm{m}(x) = \begin{cases}
    1, & \text{if} \quad p\left(x\right) > \theta \quad  \text{(source classified as a GC)}\,,\\
    0, & \text{else} \quad \text{(source not classified as a GC)}\,,
  \end{cases}
\end{equation}
with threshold $\theta$ (e.g., $\theta = 0.5$).
For models that work directly with the image data, $\xI \in \mathbb{R}^{d \times M \times M}$ denotes instead the $M \times M$ pixel-sized image with $d$ filter bands of source $i$ in our training data set, while $\yI \in \{0,1\}$ is again the corresponding label.
In this case, $p: \mathbb{R}^{d \times M \times M} \to \mathbb{R}$ maps from the image space into the real space, $x \in \mathbb{R}^{d \times M \times M}$ and $y_\mathrm{m}: \mathbb{R}^{d \times M \times M} \to \{0,1\}$.

\subsection{GC detection using tabular data}\label{sec:tabularModels}

For the task of detecting GCs using tabular feature data, we investigate logistic regression, support vector machines, neural networks, tree-based algorithms and nearest neighbour classifiers.

\subsubsection{Logistic regression}

Logistic regression is a linear classifier, which means that the decision rule for identifying GCs depends linearly on the input $x$.
The decision rule of logistic regression can be visualised as a hyperplane with normal vector $W$ (weights) and offset $b$ (biases) in the input space that separates the two classes.
This yields a high degree of interpretability, since the learned weights $W$ directly tell us which features in $x$ are important for classifying a source as a GC.
If such a separation via a hyperplane is not possible, the data is not linearly separable and more expressive models like support vector machines, neural networks, and trees are required to achieve better results.

\subsubsection{Support vector machines}

Similar to logistic regression, linear support vector machines \citep{boser1992training} use a hyperplane $W$ to distinguish classes, but the hyperplane is chosen such that the distance to data points close to it is maximised.
To deal with non-linearly separable data, the kernel trick \citep{aizerman1964theoretical} is commonly applied: instead of using the features $x$ directly, new features are obtained by applying a kernel function $\kappa(\cdot, \cdot)$ to $x$ and the training samples $
\xI$, $\kappa\left(\xI,\, x\right)$.
A prominent choice for the kernel is the Gaussian radial basis function $\kappa\left(\xI,\, x\right) = \exp\left(-\gamma\, \|\xI - x\|^2\right)$ with hyperparameter $\gamma \in \mathbb{R}_+$.
This is equivalent to mapping the features $x$ into a high-dimensional vector space where a linear separation of classes is easier.

\subsubsection{Neural networks}\label{sec:neuralnet}

Neural networks extend the previous methodologies to deep architectures, where an input is processed by sequentially applying the following operations multiple times: (i) linearly map the input (as in logistic regression), (ii) apply a non-linear activation function $\phi$, and (iii) use the output as the new input.
By training this hierarchy of functional modules (also called layers) using the error backpropagation algorithm \citep{linnainmaa1970representation,werbos1982applications,Rumelhart1986}, neural networks find a suitable non-linear transformation $F(x)$ of the input features to linearly separate the underlying classes.
In fact, most often the last layer of a neural network is logistic regression applied on the transformed feature vectors $F(x)$ (see \cref{app:neuralnet}).
In the special case of linear activation functions ($\phi(x) = x$), neural networks become identical to logistic regression.

Although deep neural networks often reach outstanding performance levels \citep{lecun2015deep}, they suffer from a lack of interpretability \citep{arrieta2020explainable,linardatos2021explainable}.
Several approaches have been proposed to explain the output of a neural network retrospectively.
Model-agnostic methods, such as LIME \citep{ribeiro2016should}, train linear surrogate models that behave like the parent model locally (around a single feature vector) to generate explanations.
Alternatively, one can identify which changes to the input feature vector $x$ mostly affect the classification output of the model.
This information can, for instance, be obtained from the model gradient $\nabla_x p(x)$ \citep{sundararajan2017axiomatic,montavon2017explaining}.
In addition, neural network architectures like TabNet\footnote{For our data, without thoroughly optimising all available hyperparameters, TabNet reaches similar performance levels (but higher false positive rates) than standard deep neural networks (not shown).} \citep{arik2020tabnet} and BagNet \citep{brendel2019approximating} have recently been proposed that are more interpretable by design and provide insight into which features the network used for its prediction.
To further increase the transparency of neural networks, techniques like Monte Carlo Dropout \citep{gal2016dropout} can be used to provide both a class prediction and a sensible estimate of the neural network's uncertainty (see \cref{app:neuralnet,app:dropout}).

\subsubsection{Decision trees and forests}

A decision tree \citep{cart84} is a tree-structured series of questions that are used to figure out which class an input $x$ belongs to.
It is created by repeatedly splitting the training data into subsets (branches) by finding appropriate conditions for elements of the feature vector $x$.
The reasons why a source is assigned a certain class is given by the decision path in the tree and is hence completely comprehensible.

However, decision trees are prone to overfitting the training data.
This downside can be alleviated by building ensembles of decision trees, so-called RFs \citep{breiman2001random}, where each tree is grown from a random sub-selection of features, that is, elements in $x$ on which conditions can be applied (called feature bagging) and training data samples (called bootstrapping).
In this case, the final classification is given by the majority vote of all trees in the ensemble.
In a RF, all trees are created independently.
Instead, one can use a technique called gradient boosting, where trees are sequentially added to the ensemble and subsequent trees learn to correct the mistakes of their predecessors.

Both for RFs and boosted trees, the decision pathways can be investigated -- although less conveniently than for a single decision tree.
A more global view of the inner workings of RFs and boosted trees can be obtained by looking at which features have been most important for building the decision trees of the ensemble.
Moreover, by using model-agnostic methods like LIME, explanations for individual classifications can be generated as well.

\subsubsection{k nearest neighbours}

A k nearest neighbour (kNN) classifier searches for the k sources in the training data that are closest, for example, with respect to the Euclidean norm, to the source we want to classify and assigns it the label that the majority of the k nearest neighbours have.
Thus, different from the previous methods, no parameters or decision rules are learned from the training data, but the training data itself is used during inference time\footnote{However, different features might cover different value ranges (or even have different units), which have to be rescaled accordingly for kNN to produce good results.}.
One benefit of this method is that we can directly look at the training samples (and their distances to $x$) that were used to make the class prediction for $x$.
However, a downside is that the inference time grows with the training set size, making it often slower than alternative methods like neural networks and RFs.

\subsection{GC detection using image data}

For the task of detecting GCs from image data, we use again the kNN and RF algorithms from the previous section as baselines.
In addition, we also apply CNNs \citep{fukushima1982neocognitron,lecun1998gradient}, which are currently one of the most prominent models for image recognition tasks \citep{lecun2015deep}.

Similar to a normal neural network (\cref{sec:neuralnet}), a CNN is a hierarchy of functional units, with the main difference being that in initial layers, linear maps take the form of convolutions with learnable filters (weights) over the spatial structure of the input.
Like classical neural networks, CNNs suffer from a lack of interpretability. Although gradient and surrogate methods like LIME could be used to identify which pixels of the input image mostly contributed to the model's prediction, this only leads to limited insights due to the low spatial resolution of GC cut-out images.
For this reason, we further investigate a variant of the CNN architecture that replaces the logistic regression layer at the end of the CNN architecture with a method that is more interpretable by default \citep{papernot2018deep} -- in this case, kNN.
This works as follows: instead of using the image data for the kNN search, we project training images $\xI$ and the test image $x$ into the latent space of the CNN, $\xI \mapsto F(\xI)$ and $x \mapsto F(x)$, and use these representations for the kNN algorithm\footnote{For a detailed description of $F(x)$, see \cref{app:CNN}.}.
This allows us to investigate the k nearest neighbours used to predict the class label in the original image space (or even the tabular feature space), increasing the transparency of the algorithm without impairing performance.

Again, as in \cref{sec:neuralnet}, Monte Carlo Dropout can be used during inference time to provide both a class prediction and an uncertainty estimate from CNNs (see \cref{app:dropout}).

\subsection{Data splits and rescaling}\label{sec:data_preprocessing}

We look at two specific cases that mimic how we would realistically approach the problem of identifying GCs from photometric surveys:
\begin{enumerate}
    \item The data is available as a collection of sources from different clusters, with a fraction of sources being labelled. 
    In this case, we merge the Fornax and Virgo data set. The resulting data set is randomly split in a 76:4:20 ratio into training, validation and test data.
    For cross-validation, we repeat this process ten times to generate different splits.
    \item Training data is available only for one cluster (e.g., Virgo) and the trained model is used to find GCs in the galaxies of  another cluster (e.g., Fornax).
    In this case, training and validation data is generated by randomly splitting the sources of, for example, Virgo in a 95:5 ratio, while the sources of Fornax are only used for testing.
\end{enumerate}
The validation data is used to tune hyperparameters and select the best model, while the test data is only used to measure the performance of the final model.

We further rescale all tabular features such that they span similar value ranges\footnote{This is not required for tree-based algorithms, but strongly improves the performance of algorithms like neural networks and kNN.}.
The general formula for rescaling a feature $\xI_k$ is given by
\begin{equation}
    \xI_k \mapsto \frac{\xI_k - \alpha_k}{\beta_k} \,.
\end{equation}
The scaling parameters $\alpha_k$ and $\beta_k$ are calculated from the training split and then applied to the training, validation and test data.
Details on how $\alpha_k$ and $\beta_k$ have been chosen for individual features $k$ can be found in Appendix \ref{app:rescaling}.
The image data are not rescaled.

In the training data, sources that contain 'NaN'\footnote{Not a Number \citep{NaN}.} tabular feature values are dropped during training.
For the validation and test data, 'NaN' entries are replaced by the median value of the corresponding tabular feature obtained from the training split.
Some of the tabular features were dropped from the data altogether, for example, non-meaningful features like sky coordinates or features with a high fraction of 'NaN' values.
'NaN' values in the images are replaced by 0.

\subsection{Evaluation metrics}\label{sec:metrics}
\begin{table*}[!htb]\renewcommand{\arraystretch}{1.2}
\caption{Classifying GCs from the Virgo and Fornax tabular data sets. }\vspace{-4mm}
\begin{center}
\begin{tabular}{c | c c c c}
\hline\hline
Method & TPR & FPR & FDR & AUC ROC \\
\hline\hline
Logistic Regression & \me{0.761}{0.007} & \me{0.054}{0.002} & \me{0.204}{0.005} & \me{0.951}{0.001} \\
Support Vector Machine (linear) & \me{0.778}{0.007} & \me{0.053}{0.002} & \me{0.196}{0.005} & --  \\
Support Vector Machine (radial) & \me{0.858}{0.005} & \me{0.042}{0.002} & \me{0.151}{0.006} & -- \\
Nearest Neighbour & \me{0.897}{0.005} & \me{0.040}{0.002} & \me{0.139}{0.007} & -- \\
12 Nearest Neighbours  & \me{0.914}{0.005} & \me{0.030}{0.001} & \me{0.100}{0.004} & \me{0.987}{0.001}\\
Decision Tree & \me{0.892}{0.009} & \me{0.036}{0.003} & \me{0.127}{0.010} & \me{0.983}{0.001} \\ 
Random Forest & \me{0.909}{0.005} & \me{\best{0.018}}{0.001} & \me{0.066}{0.005} & \me{\best{0.992}}{0.001} \\
AdaBoost & \me{0.910}{0.005} & \me{\best{0.018}}{0.001} & \me{\best{0.065}}{0.005} & \me{\best{0.992}}{0.001}  \\
CatBoost  & \me{0.920}{0.006} & \me{0.025}{0.002} & \me{0.089}{0.007} & \me{0.990}{0.001}\\
Neural Network (29-1) & \me{0.756}{0.007} & \me{0.056}{0.002} & \me{0.210}{0.005} & \me{0.948}{0.001} \\
Neural Network (29-100-100-1) & \me{\best{0.932}}{0.009} & \me{0.028}{0.003} & \me{0.097}{0.010} & \me{\best{0.992}}{0.001} \\
\hline
\end{tabular}
\label{tab:featureLearningALL}
\end{center}
\tablefoot{
The reported results are averages over 10 random splits (train, validation and test). Uncertainties are given as standard deviations and the used decision threshold is $0.5$. Support vector machines and nearest neighbour only provide a class label and hence, no AUC ROC is reported.
For neural networks, the number of neurons per layer is given in brackets.
}
\end{table*}

\begin{table*}[!htb]\renewcommand{\arraystretch}{1.2}
\caption{Training on tabular data from Virgo and evaluating on Fornax data.}\vspace{-4mm}
\begin{center}
\begin{tabular}{c | c c c c | c c c c | c c}
\multicolumn{1}{c}{}&\multicolumn{4}{c}{averaged per galaxy}&\multicolumn{6}{c}{average over all sources}\\
\hline\hline
Method & TPR & FPR & FDR & AUC ROC & TPR & FPR & FDR & AUC ROC & \# TPs & \# FPs \\
\hline\hline
Logistic Regression & 0.63 & 0.05 & 0.33 & 0.93 & 0.70 & 0.05 & 0.16 & 0.94 &	4331 & 832 \\
Support Vector Machine (linear) & 0.66 & 0.05 & 0.31 &  -- & 0.73 & 0.05 & 0.15 & -- &	4474 & 793 \\
Support Vector Machine (radial) & 0.73 & 0.04 & 0.23 & -- & 0.81 & 0.04 & 0.12 & -- & 4993 & 656 \\
Nearest Neighbour & 0.86 & 0.05 & 0.26 & -- 
& 0.90 & 0.06 & 0.14 & --
& 5536 & 915 \\
12 Nearest Neighbours & 0.86 & 0.04 & 0.19 & 0.98 & 0.91 & 0.04 & 0.11 & 0.98 & 5597 & 678 \\
Decision Tree & 0.87 & 0.05 & 0.24 & 0.97 & 0.89 & 0.05 & 0.13 & 0.98 & 5496 & 803 \\ 
Random Forest & 0.86 & \best{0.03} & \best{0.14} & \best{0.99} & 0.91 & \best{0.03} & \best{0.08} & \best{0.99} & 5602 & \best{470} \\
AdaBoost & 0.86 & \best{0.03} & 0.15 & \best{0.99} & 0.91 & \best{0.03} & \best{0.08} & \best{0.99} & 5596 & 473 \\
CatBoost & \best{0.89} & 0.04 & 0.19 & 0.98 & \best{0.93} & 0.04 & 0.10 & \best{0.99} & 5709 & 656 \\
Neural Network (29-1) & 0.61 & 0.05 & 0.33 & 0.92 & 0.68 & 0.05 & 0.16 & 0.93 & 4159 & 801 \\
Neural Network (29-100-100-1) & \best{0.89} & 0.04 & 0.21 & 0.98 & \best{0.93} & 0.05 & 0.11 & \best{0.99} & \best{5754} & 739 \\
\hline
\end{tabular}
\label{tab:featureLearningVirgo2Fornax}
\end{center}
\tablefoot{The Fornax data contains in total 21767 sources with 6161 catalogued GCs. Results are given for a decision threshold of $0.5$. Support vector machines and nearest neighbour only provide a class label and hence, no AUC ROC is reported. For neural networks, the number of neurons per layer is given in brackets.}
\end{table*}

In the following, we denote the test data set by $\mathcal{T}$.
For convenience, we introduce the following function
\begin{equation}
    \mathrm{ID}(y, y_\mathrm{m}(x), l) = \begin{cases}
    1,  & \text{if } y = y_\mathrm{m}(x) \text{ and } y = l\,,\\
    0, & \text{otherwise}\,,
  \end{cases}
\end{equation}
to test whether the model prediction agrees with the according test label.
With this set up, we can define the following quantities: the number of True Positives (\TP), True Negatives (\TN), False Positives (\FP) and False Negatives (\FN):
\begin{align}
    \TP &= \sum_{(x,y) \in \mathcal{T}} \mathrm{ID}\left(y,\, y_\mathrm{m}(x),\, 1\right)\,, \\
    \TN &= \sum_{(x,y) \in \mathcal{T}} \mathrm{ID}\left(y,\, y_\mathrm{m}(x),\, 0\right)\,, \\
    \FP &= \sum_{(x,y) \in \mathcal{T}} \mathrm{ID}\left(1-y,\, y_\mathrm{m}(x),\, 1\right)\,, \\
    \FN &= \sum_{(x,y) \in \mathcal{T}} \mathrm{ID}\left(1-y,\, y_\mathrm{m}(x),\, 0\right) \,.
\end{align}
\begin{table*}[!htb]\renewcommand{\arraystretch}{1.2}
\caption{Classifying GCs from the combined image data sets of Virgo and Fornax.}\vspace{-4mm}
\begin{center}
\begin{tabular}{c | c c c c}
\hline\hline
Method & TPR & FPR & FDR & AUC ROC \\
\hline\hline
Nearest Neighbour & \me{0.842}{0.005} & \me{0.033}{0.002} & \me{0.122}{0.006} & $-$  \\
12 Nearest Neighbours & \me{0.847}{0.008} & \me{0.023}{0.001} & \me{0.088}{0.004} & \me{0.975}{0.002} \\
Random Forest & \me{0.838}{0.005} & \me{0.026}{0.001} & \me{0.100}{0.005} & \me{0.981}{0.001}  \\
\hline
Convolutional Neural Network (CNN) & \me{\best{0.929}}{0.015} & \me{0.019}{0.003} & \me{0.068}{0.011} & \me{\best{0.994}}{0.001}  \\
CNN + Nearest Neighbour & \me{0.912}{0.006} & \me{0.026}{0.001} & \me{0.094}{0.004} & $-$  \\
CNN + 12 Nearest Neighbours & \me{0.927}{0.007} & \me{\best{0.017}}{0.001} & \me{\best{0.061}}{0.004} & \me{0.990}{0.001}  \\
\hline
\end{tabular}
\label{tab:imageLearningALL}
\end{center}
\tablefoot{The reported results are averages over 10 random splits (train, validation and test). Uncertainties are given as standard deviations and the used decision threshold is $0.5$.}
\end{table*}
\begin{table*}[!htb]\renewcommand{\arraystretch}{1.2}
\caption{Training on image data of Virgo and evaluating on Fornax image data.}\vspace{-4mm}
\begin{center}
\begin{tabular}{c | c c c c | c c c c | c c}
\multicolumn{1}{c}{}&\multicolumn{4}{c}{averaged per galaxy}&\multicolumn{6}{c}{average over all sources}\\
\hline\hline
Method & TPR & FPR & FDR & AUC ROC & TPR & FPR & FDR & AUC ROC & \# TPs & \# FPs \\
\hline\hline
Nearest Neighbour & 0.77 & 0.04 & 0.24 & $-$ & 0.81 & 0.04 & 0.12 & $-$ & 4982 & 650 \\
12 Nearest Neighbours & 0.75 & 0.03 & 0.20 & 0.96 & 0.81 & 0.03 & 0.09 & 0.97 & 4969 & \best{476} \\
Random Forest & 0.74 & 0.03 & 0.18 & 0.96 & 0.81 & 0.03 & 0.09 & 0.97 & 4985 & 495 \\
\hline
Convolutional Neural Network (CNN) & \best{0.90} & \best{0.03} & \best{0.14} & \best{0.99} & \best{0.94} & 0.04 & 0.09 & \best{0.99} & \best{5771} & 563 \\ 
CNN + Nearest Neighbour & 0.87 & 0.04 & 0.18 & $-$ & 0.91 & 0.04 & 0.10 & $-$ & 5634 & 633  \\ 
CNN + 12 Nearest Neighbours & 0.89 & \best{0.03} & \best{0.14} & 0.98 & 0.93 & \best{0.03} & \best{0.08} & 0.98 & 5729 & 517   \\ 
\hline
\end{tabular}
\label{tab:imageLearningVirgo2Fornax}
\end{center}
\tablefoot{The Fornax data contains in total 21767 sources with 6161 catalogued GCs. Results are given for a decision threshold of $0.5$.}
\end{table*}

In other words, TP and TN are the amount of actual GCs and non-GCs found by the model, respectively, while FP and FN are the number of wrongly detected GCs and non-GCs, respectively.

To evaluate our model, we are interested in identifying how reliably GCs are found, but also how many GCs are erroneously proposed by the model.
To do this, we use the True Positive Rate (\TPR) and False Positive Rate (\FPR)
\begin{align}
  \mathrm{TPR} &= \frac{\TP}{\TP + \FN}\,, \\
  \mathrm{FPR} &= \frac{\FP}{\FP + \TN}\,.
\end{align}
Thus, the TPR is the fraction of GCs found from the test data set, while the FPR is the fraction of non-GCs that have been misclassified as a GC.
In our case, the goal is to detect as many real GCs as possible, meaning that we strive for a TPR close to 1 and a FPR close to 0.

Since our data set often contains many more sources that are non-GCs, it might still be that a large fraction of detected GCs are false positives even if the TPR and FPR values are very good.
For this reason, we also look at the False Discovery Rate (\FDR)
\begin{equation}
    \FDR = \frac{\FP}{\FP + \TP} \,,
\end{equation}
which tells us how many of the GCs the model predicts are actually false positives.

Most of the used models provide a scalar output for each source from which the class label is inferred using a threshold $\theta$.
For instance, some models return the probability of a source being a GC and the classification result is obtained by comparing if it exceeds $\theta$, for instance $\theta = 0.5$.
Through $\theta$ the trade-off between TPR and FPR can be adjusted.
By increasing $\theta$, only sources where the model is more confident will be classified as GCs, reducing the amount of FPs (but also possibly reducing the number of TPs).
How the TPR changes with respect to the FPR is quantified by the Receiver Operating Characteristic (ROC) curve, which plots the FPR against the TPR for varying thresholds.
The ROC curve is bounded by $\TPR = 0$, $\FPR = 0$ (all sources are classified as non-GCs) and $\TPR = 1$, $\FPR = 1$ (all sources are classified as GCs).
How a model responds to changing the threshold can be quantified via the area under the ROC curve (AUC ROC), with values closer to 1 being better, meaning that the threshold can be set in a way that even for low FPRs, high TPRs are achieved.

\subsection{Software implementation} \label{sec:software}
We implemented the aforementioned models in Python 3.8.8 using the Scikit-Learn 0.24.2 \citep{scikit-learn}, PyTorch 1.10.0 \citep{NEURIPS2019_9015}, and \mbox{CatBoost 1.0.3 \citep{NEURIPS2018_14491b75}} libraries.
Simulations were performed on a commercial CPU (AMD Ryzen 9 5900HS@3.3GHz) and GPU (NVIDIA GeForce RTX 3060 Laptop GPU).
The data sets and code with hyperparameters are publicly available on github\footnote{\url{https://github.com/dodo47/GCDetection}}.

\section{Results}\label{sect:results}

\subsection{Globular cluster detection}
\begin{figure}
    \centering
    \includegraphics[width=0.925\columnwidth]{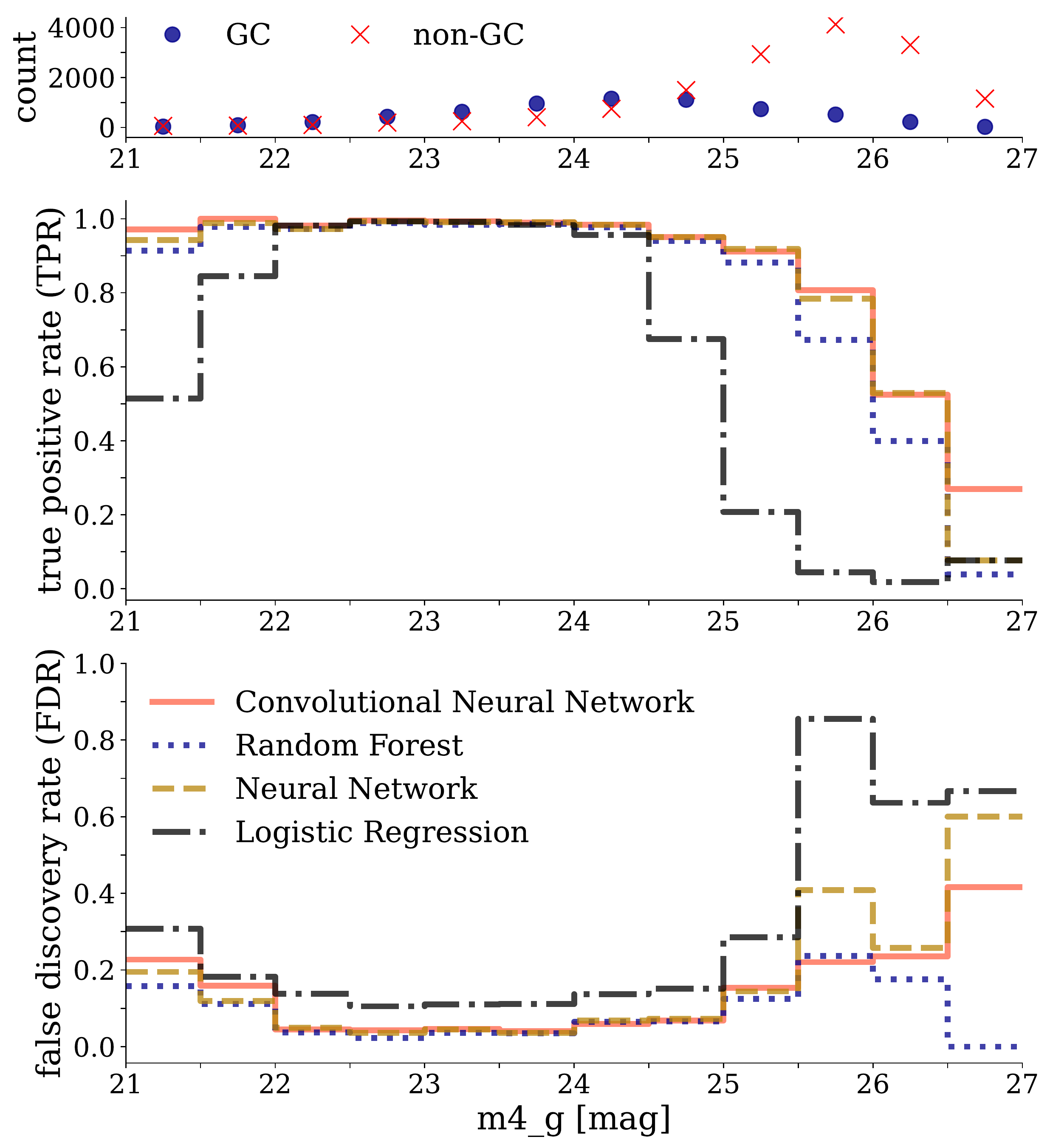}
    \caption{Model performance for sources of different magnitude ranges, separated into bins of size $0.5$ mag, for the case of training on data of the Virgo cluster and evaluating on data of Fornax. At the top, the number of GCs and non-GCs per bin are shown.}
    \label{fig:magnitude_dependence}
\end{figure}

\begin{figure*}
    \centering
    \includegraphics[width=0.8\textwidth]{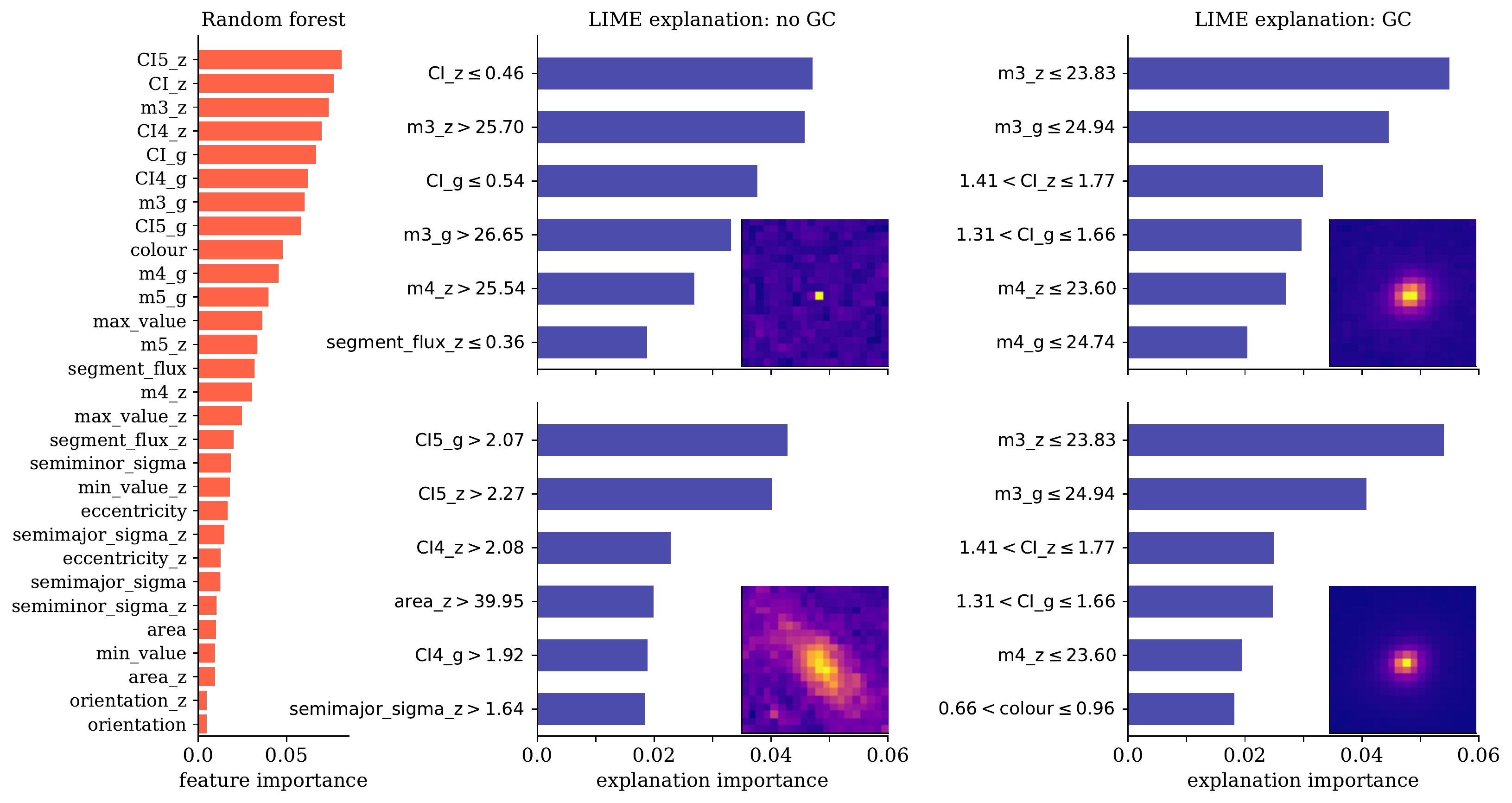}
    \caption{Feature importance of a RF trained on Virgo data (left) and explanations for individual RF predictions generated with LIME (right).}
    \label{fig:feature_importance}
\end{figure*}

\begin{figure*}
    \centering
    \includegraphics[width=0.7\textwidth]{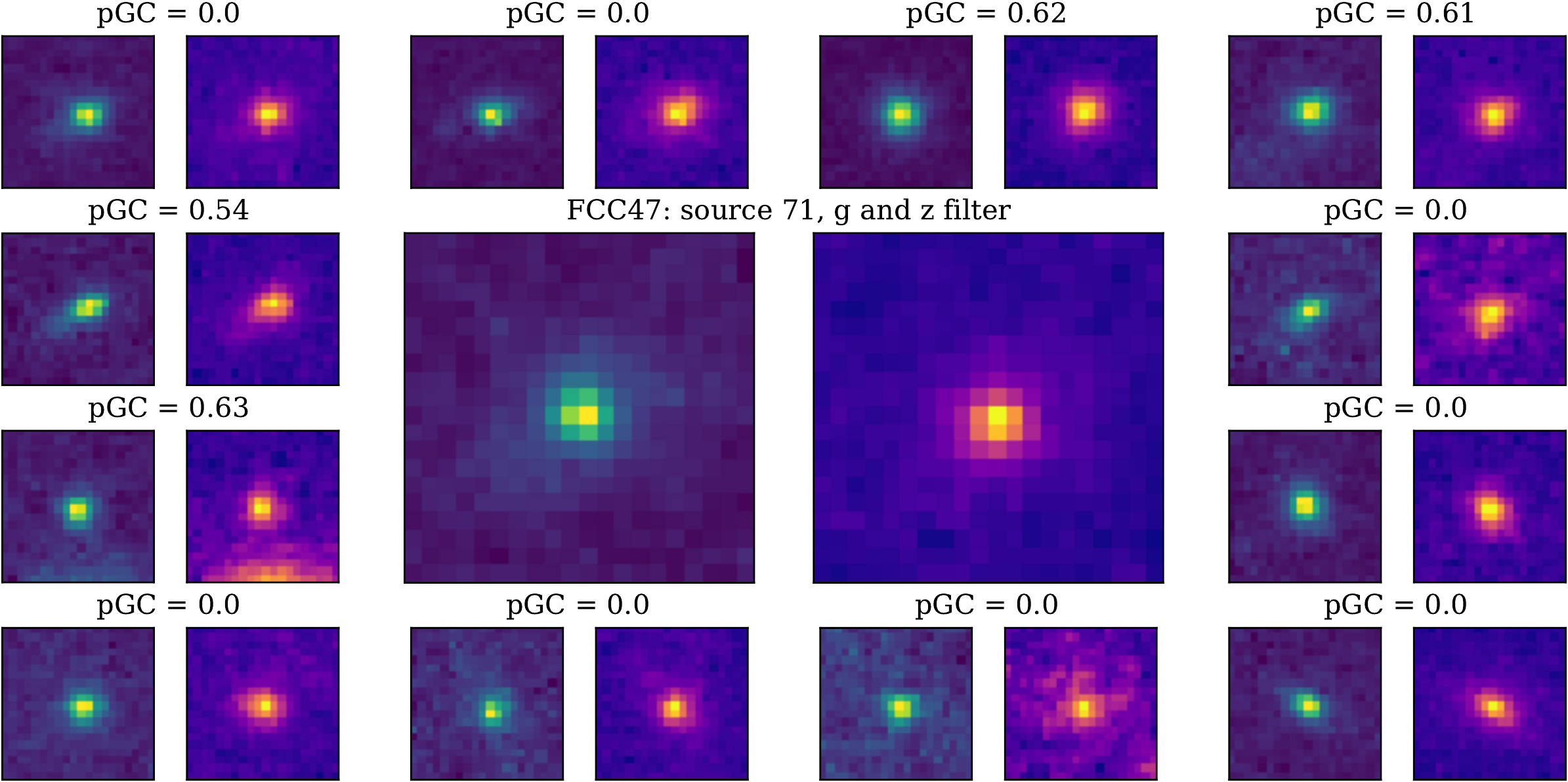}
    \caption{Images of the k nearest neighbours with labels (surrounding) used to classify source 71 of FCC47 (centre). The nearest neighbours are not found in the image space $x$, but in the latent space of the CNN $F(x)$.}
    \label{fig:CNNexplanation}
\end{figure*}

We applied the described models of \cref{sec:tabularModels} on the tabular feature data, once for the scenario of (i) predicting GCs from a mixture of sources (\cref{tab:featureLearningALL}) and (ii) for training the network on data from the Virgo cluster to find GCs in the Fornax cluster (\cref{tab:featureLearningVirgo2Fornax}).
As a cross-reference to the boosted tree implementation of Scikit-Learn (AdaBoost), we also deployed CatBoost developed by Yandex, which promises good performance even with the provided default hyperparameters \citep{NEURIPS2018_14491b75}.

Less complex models like logistic regression perform not optimally in both scenarios, missing around 25-40\% of detectable GCs while producing a high number of false positives (FPR $\sim 5$\% and FDR $\sim 16 - 20$\%).
The best performance is reached by tree-based algorithms like RFs and boosted trees as well as deep neural networks, that is, neural networks with more than two layers.
Especially RFs and boosted trees reach a high true positive rate of around 90\% while producing the lowest number of false positives (FPR $\sim 2-$3\% and FDR $\sim 6-$8\%).
For instance, in the case of predicting GCs in Fornax, the number of false positives is almost halved compared to logistic regression.
In contrast, deep neural networks are capable of finding slightly more GCs (TPR $\sim 90-$94\%), but with the downside of also producing more false positives (FPR $\sim 2-$4\% and FDR $\sim 10$\%).

The obtained results for both scenarios are consistent, with training on sources from mixed clusters leading to slightly better performance levels.
This is to be expected because the data of both surveys were not obtained under identical intrinsic conditions, as Fornax is $\sim 5$ Mpc more distant and hence completeness levels are different. The same GC will appear fainter and with a slightly smaller angular size in Fornax than in Virgo. 
However, it is promising that a model trained on data from one survey can be used to detect GCs in another one, given that the data statistics agree sufficiently. 

The second evaluation scenario allows us to evaluate the test performance of each model per Fornax galaxy (\cref{tab:featureLearningVirgo2Fornax}, left).
This mostly leads to an increase of the FDR as some low-mass galaxies only contain a small number of GCs.
Hence, the model also only detects a small number of GCs, of which a larger fraction are false positives even if the FPR is rather low.

On the image data, both kNN and RFs suffer from a drop in performance, which is to be expected since both models are known to perform best on tabular data (\cref{tab:imageLearningALL,tab:imageLearningVirgo2Fornax}).
However, both CNNs and CNNs combined with kNN as a last layer perform well in both scenarios, reaching similar or even slightly better performance levels than tree-based methods on the tabular data.
Thus, both CNN-based methods on the image data as well as tree-based methods on the tabular data reach good performance levels, that is, high TPR and low FPR and FDR, on both of the investigated scenarios.
Similar results are obtained when training models on data of the Fornax cluster and testing on Virgo (see Appendix, \cref{tab:featureLearningFornax2Virgo,tab:imageLearningFornax2Virgo}).
ROC curves for several models are shown in \cref{app:roc}.

In both cases, model performance strongly depends on the magnitude range of the sources, as Fig. \ref{fig:magnitude_dependence} shows.
For example, for sources with intermediate values of m4\_g (22 mag $<$ m4\_g $\leq$ 24.5 mag), RFs trained on tabular data reach a TPR of $\sim 97.22-98.83$\% and a FDR of $\sim 2.31-6.48$\% while CNNs trained on image data reach a TPR of $\sim 98.15-99.53$\% and a FDR of $\sim 4.05-5.90$\%.
For fainter sources (m4\_g $> 24.5$ mag), TPR and FDR become worse but remain above $\sim$90\% and below $\sim$15\%, respectively, for magnitudes of up to $25.5$ mag.
Especially compared to logistic regression -- which also only reaches a FDR of $\sim 10.5-13.8$\% in the magnitude range $22 - 24.5$ mag -- more complex models are capable of generalising towards sources with much lower or higher magnitudes.
For example, CNNs still reach a TPR of $\sim$81\% and a FDR of $\sim$22\% for sources in the magnitude range of $25.5 - 26$ mag, as well as $\sim$52\% and $\sim$24\%, respectively, for the range of $26 - 26.5$ mag, but in general the performance drops in magnitude ranges where only few sources are present in the input data.
Similar results are observed when training on data from Virgo and evaluating on data from Fornax (see \cref{app:magnitude}).

\subsection{Model interpretability}

Although model performance is an important indicator, it is by far not the only criteria that should be considered when investigating the usability of a model \citep{arrieta2020explainable,linardatos2021explainable}.
Especially when the results of a machine learning model are utilised in subsequent projects, it is important that models also provide a degree of interpretability that allows proper investigation of the model's decision process and the data used to train the model.
Although an extensive evaluation of model interpretability in the form of a survey is out of scope for this work, we demonstrate for selected examples how more insight into the inner workings of machine learning models can be obtained.

In case of tree-based algorithms like RFs, a first overview of what the model focuses on can be acquired by looking at which features have been most important in building the decision trees of the ensemble.
For a RF trained on tabular data of Virgo, we can see that, as expected, features such as orientation (position angle of the source) are least important, while concentration indices and magnitudes are most important (\cref{fig:feature_importance}, left).

Explanations for individual classifications $\left(x, y_\mathrm{m}(x)\right)$ can be obtained using the model-agnostic LIME method.
In LIME, a linear surrogate model is trained that approximates the behaviour of the RF in the local neighbourhood of the input $x$. 
This is achieved by randomly generating inputs $x^{(i)}_x$ around $x$, evaluating them using the RF to get its prediction $y_\mathrm{m}(x^{(i)}_x)$ and using the $\left(x^{(i)}_x, y_\mathrm{m}(x^{(i)}_\mathrm{x})\right)$ tuples as the training set for the surrogate model.
To get more expressive explanations, the features of $x^{(i)}_x$ are further discretised into value ranges.
Since the surrogate model is linear, the feature ranges with the highest absolute weights are then used as an explanation for the original model's decision.
To guarantee that the created explanations are somewhat reliable, we use the statistical stability indices introduced in \cite{visani2020statistical} for LIME. 

A demonstration of LIME is shown in the right panels of \cref{fig:feature_importance} for four different sources -- two which are classified as non-GCs and two which are classified as GCs by the RF.
For the two non-GCs, we picked examples where the images can be used to easily evaluate explanations provided by LIME.
In the top example, LIME states that the concentration indices and magnitudes are too low, meaning that the source is not sufficiently extended and too faint to be a GC.
In the bottom example, LIME states the opposite: the large concentration indices as well as the large area covered by the source are the reasons why it is not classified as a GC.
At least in these straightforward cases, the explanations provided by LIME are consistent with what one would expect from looking at the images of the sources.
For the two sources classified as GCs, the LIME explanation focuses on different features, in particular the $g$ and $z$-band magnitudes and the concentration indices. This is encouraging since magnitudes (and consequently colours) as well as concentration indices are traditionally used to identify star clusters in photometric data (e.g. \citealt{Amorisco2018, Adamo2020, Thilker2022}).
Moreover, for both GCs, the provided explanations are very similar, showcasing a certain degree of consistency of LIME.
This is also true for explanations generated for other GCs (see \cref{app:limeExp}).

In case of the CNN with kNN layer, the images of the training sources that have been used for the prediction can be directly investigated together with their labels and distances to the input image $x \in \mathbb{R}^{d\times M \times M}$ in the latent space $F(\cdot)$ of the CNN.
For instance, in \cref{fig:CNNexplanation} we show source 71 of the Fornax galaxy FCC\,47 which has been falsely classified as a non-GC.
The k nearest neighbours show that similar images exist in the training data which are either classified as non-GCs or only classified as GCs with a low confidence.
We can further see that many of these sources show an elongated structure around the central source which is also slightly present in source 71 of FCC47 and might be the reason why it has not been classified as a GC.

\section{Discussion}
\label{sect:discussion}

\subsection{False negatives and false positives}
We first explore the properties of sources that are either false negatives or false positives -- even in the best performing models -- to understand whether these misclassified sources are informative about inherent problems in the modelling.
\begin{figure}[t!]
    \centering
    \includegraphics[width=0.45\textwidth]{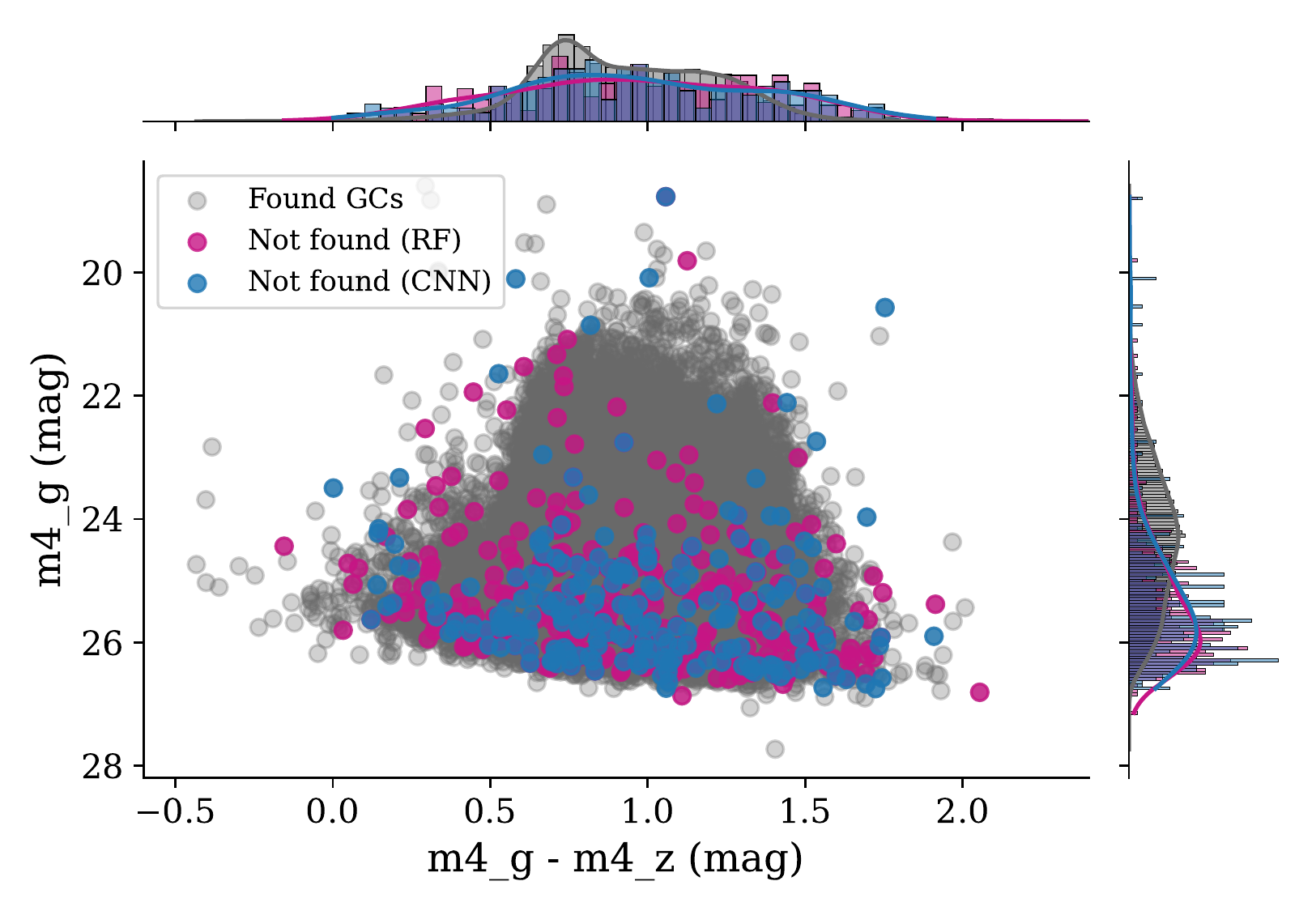}
    \caption{Colour-magnitude diagram showing the recovered GCs in grey. Sources in pink and blue refer to sources labelled as GCs that were not found (false negatives) by the RF or CNN method applied to the full data set, respectively.}
    \label{fig:not_found_CMD}
\end{figure}

Figure \ref{fig:not_found_CMD} shows a colour-magnitude diagram of all sources labelled as GCs in the original data. Sources that were not found by the RFs (applied to the tabular data) or CNNs (applied to the image data) are highlighted in colour. This figure illustrates that the false negatives tend to be faint sources, many of them with m4\_g $>$ 26 mag, much fainter than the bulk of GCs. The faintness of these sources could be the reason for the misclassification, because identification of fainter sources is generally more difficult due to relatively higher backgrounds and noise levels. 

\begin{figure}[ht]
    \centering
    \includegraphics[width=0.45\textwidth]{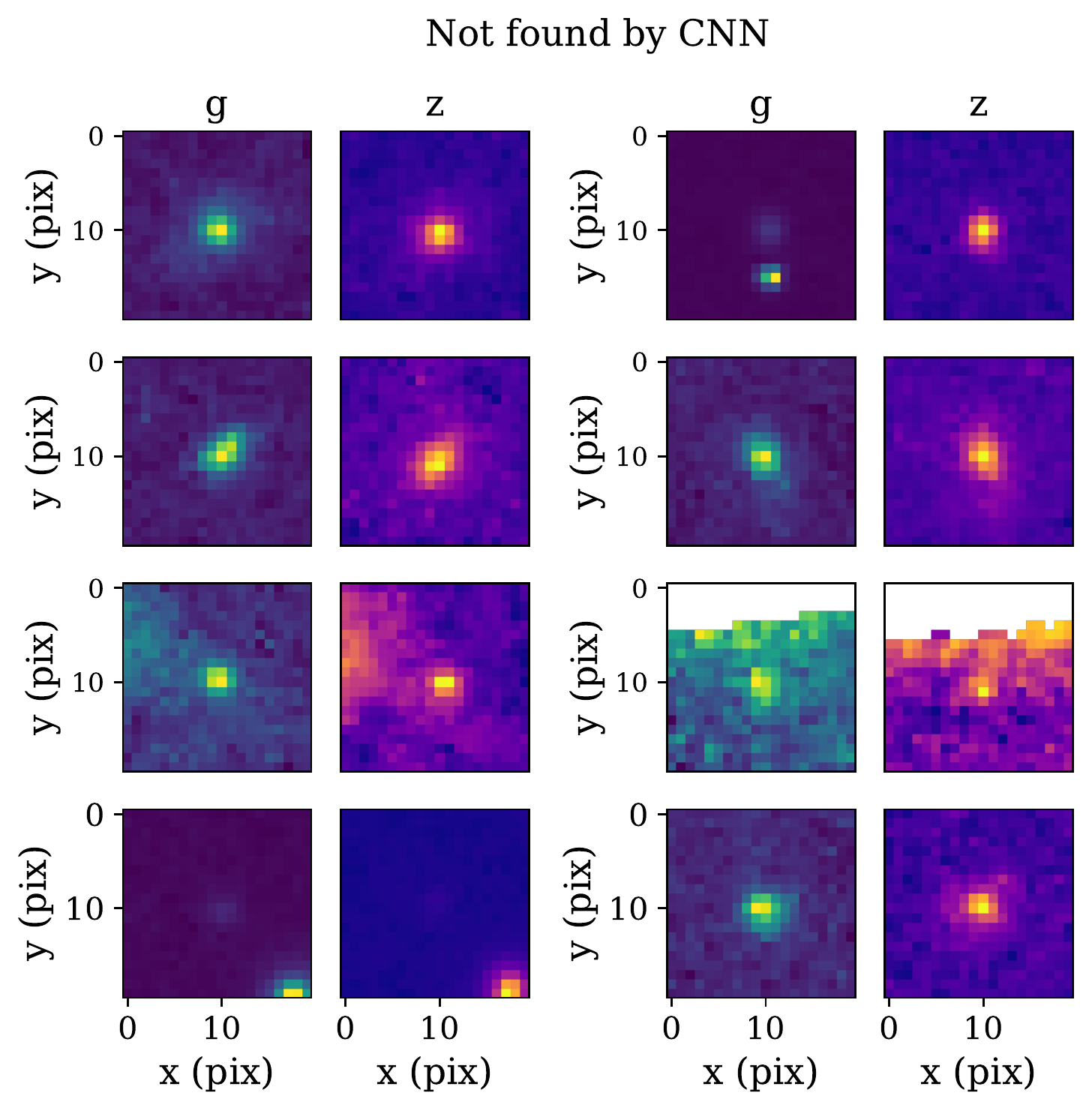}
    \caption{Examples of sources that are labelled as GCs, but were not found by the CNN on the image data (false negatives). While for some sources, the reason for the non-detection is not immediately obvious, others show extended backgrounds or are located close to the detector edges.}
    \label{fig:not_found_CNN}
\end{figure}

Exploring the false negatives further, Fig. \ref{fig:not_found_CNN} shows the $g$ and $z$-band images of a few GCs that were not detected by the CNN. For some sources, reasons for the failure of detection can be identified from these images, such as image artefacts in one filter, extended backgrounds, or proximity to detector edges. However, for other sources the reasons cannot be inferred from the images by eye. For those, interpretable methods as presented above might hold additional information, which is illustrated in \cref{fig:CNNexplanation} for source 71 of FCC47 (top left source in \cref{fig:not_found_CNN}). 

\begin{figure}
    \centering
    \includegraphics[width=0.425\textwidth]{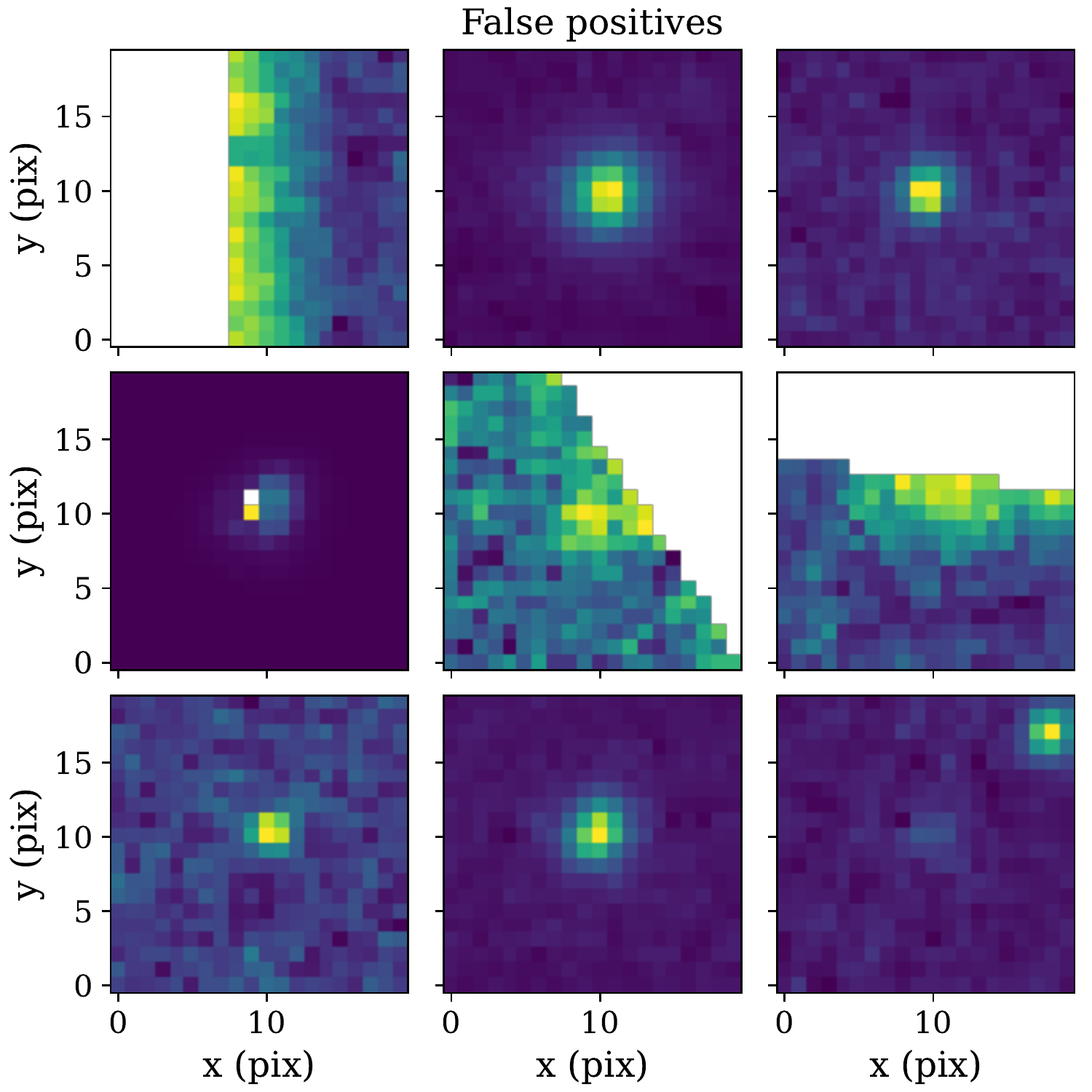}
    \caption{Examples of false positives identified with the RF method. While some sources are indistinguishable from GCs by eye, also image artefacts at the detector edges are among the false positives.}
    \label{fig:false_positives_examples}
\end{figure}

\begin{figure}
    \centering
    \includegraphics[width=0.45\textwidth]{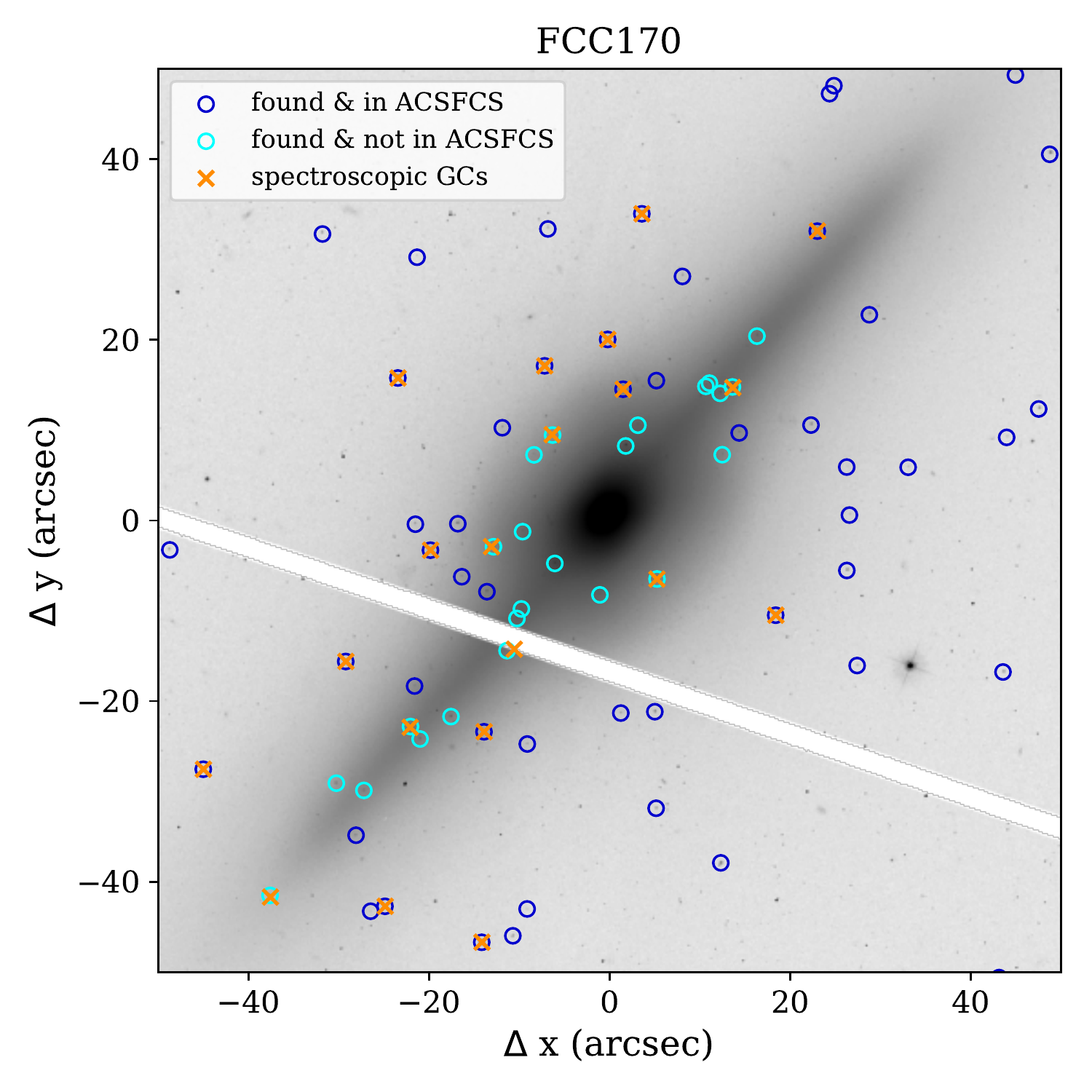}
    \caption{Zoom into the central region of the HST ACS F475W image of FCC170. Sources in blue are GCs from the ACSFCS catalogue \citep{Jordan2015} while sources in cyan were identified as GCs by the RF, but are not labelled as such in the original ACSFCS catalogue (false positives). The crosses mark spectroscopically confirmed GCs from \cite{Fahrion2020b}.}
    \label{fig:false_positives_FCC170}
\end{figure}

Similarly, exploring the images of false positives only provides in some cases insights into why a source was misclassified. Fig. \ref{fig:false_positives_examples} shows examples of several false positives produced by the RF. While clearly some of the sources are image artefacts, for example at the detector edges, others are indistinguishable from GCs by eye.

However, we have reasons to believe that at least some of these false positives are in fact true GCs which were not labelled as such in the ACSFCS and ACSVCS catalogues. 
To illustrate this, Fig. \ref{fig:false_positives_FCC170} shows a cut-out of the ACSFCS image of the edge-on S0 galaxy FCC\,170. According to the RF method trained on Virgo galaxies, this galaxy has 25 false positives and as the figure shows, 24 of them are in the central high surface brightness region of the galaxy. This region was not included in the ACSFCS catalogue, likely because subtracting a surface brightness model of this galaxy is challenging due to the X-shaped bulge and the thin disk (e.g. \citealt{Pinna2019, Fahrion2021}). As we used a simple median background filter, we have included sources in these regions and find several sources which were classified as GCs by the RF. That at least the brightest of those are in fact GCs can be seen by cross-referencing with the sample of spectroscopically confirmed GCs from \cite{Fahrion2020b}. 

This closer inspection of false negatives and positives illustrates that already excluding sources near the detector edges in addition to the detector gap leads to a cleaner sample. Nonetheless, at these low rates, consequences for inferred properties of GC systems such as luminosity functions should be minimal, especially since the existing catalogues are not perfect either. However, we note that low-mass galaxies might be affected more due to the intrinsic lower number of GCs.

\subsection{Model performance and generalisation to new data}

The number of true and false positives can be adjusted for most models via the decision threshold, which has to be set depending on the application the model is used for. 
For example, by increasing the threshold, the fraction of false positives will decrease, although with the drawback that less of the real GCs present in the data will be found as well.

In general, the trained models reach good performance levels and are suitable for detecting GCs in photometric data.
Our results agree with a recent study evaluating explainable machine learning for extracting ultra-compact dwarfs and GCs from ground-based imaging of the Fornax cluster, which reports true positives rates of $0.89 - 0.97$ and false discovery rates of $0.04 - 0.07$ \citep{Mohammadi2022}, although on a much smaller data set of $\sim$ 7700 sources containing $\sim$ 500 GCs and for six instead of only two bands, making an exact comparison impossible.
Recently published studies on detecting star clusters from HST data using machine learning also report similar results: \citep{Perez2021} use a CNN (StarcNet) and a data set of $\sim$15000 sources from the LEGUS galaxies with observations in five bands for training, validation, and testing.
They report a TPR of $\sim$81\% and a FDR of $\sim$19\% on a cluster/non-cluster classification task, reaching a TPR of $\sim$93\% and a FDR of $\sim$7\% if testing is restricted to high-mass objects (153 sources).
\cite{Whitmore2021} 
report a TPR of $\sim$82\% for a similar cluster/non-cluster classification task for sources from five galaxies of the PHANGS-HST sample using two popular CNN architectures (Resnet18 and VGG19-BN) trained via transfer learning on $\sim$5500 sources of 10 LEGUS galaxies.
When only applying their trained models to isolated objects and for detecting compact and symmetric star clusters, they reach a TPR of $\sim$92\%.
Again, due to the differences in the investigated data sets and evaluation schemes, an exact comparison with our results is difficult.

It is especially encouraging that models trained on sources of one cluster generalise to other clusters (e.g., training on Virgo to find GCs in Fornax).
However, this should be taken with a grain of salt, since generalisability is not guaranteed when applying the trained model on data that lies outside of the training domain. For instance, applying our models to NGC\,1427a, a star-forming dwarf galaxy in Fornax \citep{Mora2015, LeeWaddell2018}, leads to many false detections of sources that likely are not GCs but rather young star clusters and stellar associations. 

Both the models trained on image and tabular data reach similar maximum performance levels, most likely due to noise and imperfections both in the recorded data as well as the ground-truth labels used to train and evaluate the models.
Still, the performance might be further improved by enriching the data, for example by including more filter bands, by increasing the amount of labelled data for training, or by fine-tuning models by providing more class labels than just non-GC and GC. 
For the latter case, the catalogues of young star clusters created within the LEGUS or PHANGS projects  (e.g. \citealt{Whitmore2021, Perez2021}) could be used to build a comprehensive sample of GCs and young star clusters for training models. However, when including young star clusters, possibly additional features that encompass the specifics of young star clusters in comparison to mostly spherical GCs need to be included (see e.g \citealt{Whitmore2011} or \citealt{Deger2022}).
Furthermore, observed data might be extended using mock data, although great care has to be put into guaranteeing that artificially created data is representative of recorded data.

For the data volumes expected for future wide field missions such as Euclid or the Nancy Grace Roman Space Telescope, labelled data will be greatly outweighed by unlabelled data, hence potential methodologies to be explored in future work are unsupervised and semi-supervised approaches.

\subsection{Model complexity}

We identified ensemble-based tree methods on tabular data and CNN-based methods on image data as the two best-performing approaches.
Compared to CNNs, models like RFs have the advantage of being easy to set up: they require no feature rescaling of the data, contain only a small number of tunable hyperparameters, generally perform and generalise well due to being ensemble-based, and are appropriately interpretable.
In contrast, CNNs can be directly applied to images, but lack interpretability and come with many degrees of freedom that can be tuned, such as the model architecture, used activation functions, regularisation techniques and hyperparameters.
This certainly makes tree-based models the more attractive choice for the task of detecting GCs.
Nevertheless, CNNs have the potential of outperforming RFs especially for increased data complexities (e.g., more filters) due to their capability of learning to identify appropriate and task-dependent features in the images automatically.

\subsection{Model interpretability}

We illustrated the application of interpretable methods to provide more insight into model decisions and to help explore the training data set, for example, to find unexpected labels or artefacts.
Interpretable methods are important to increase the trustworthiness of models, easing the decision process for  whether model predictions can be trusted when applying them on data beyond the initial training domain \citep{arrieta2020explainable, linardatos2021explainable}.

In particular, the model-agnostic LIME method confirms that the investigated models focus on the proper tabular features for detecting GCs.
However, an extensive study of the quality of explanations provided by LIME is out of scope for this work.
Hence, there is no guarantee that for all instances, the explanations obtained via LIME are helpful or completely accurate in describing the original model.
In fact, how to evaluate the quality of generated explanations is an active field of research \citep{zhou2021evaluating}.
The quality of generated explanations could be quantified in future work via human-centred surveys or by deriving fidelity metrics that score how accurately the surrogate explanation describes the model's decision.

\section{Example science case: colour distributions}
Eventually, the goal of building a machine learning pipeline for detecting GCs is to use the obtained catalogues in scientific analyses, for example, to study galaxy properties. 
Although the overlap between model-generated and original catalogues is high, it is not perfect -- especially not on the level of individual galaxies. 
Thus, to demonstrate the applicability of model-generated
GC catalogues for scientific use-cases, we provide an example of a measure of interest -- the $g - z$ colour distribution of GCs in a host galaxy -- that can be derived from our model output and compare it to corresponding results derived from the original catalogues.

\begin{figure}[t!]
    \centering
    \includegraphics[width=0.46\textwidth]{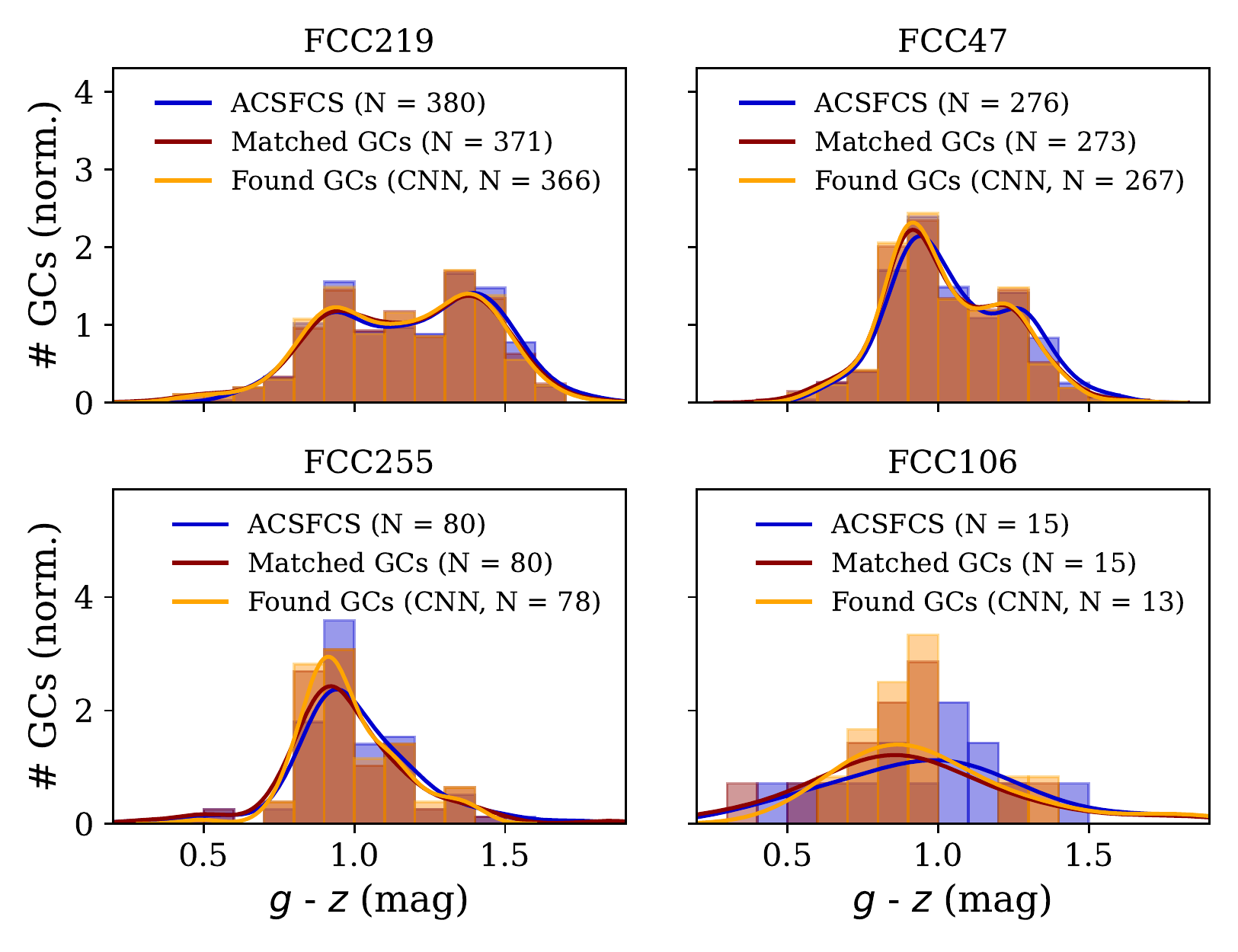}
    \caption{Four ($g$ - $z$) colour distributions of galaxies in the Fornax cluster. Blue histograms show the original ACSFCS data, red are all matched GCs in our data set and orange histograms show the sources identified as GCs by the CNN. The corresponding lines are kernel density estimations for visualisation of the histogram shapes. The legend states the number of sources in each sample.}
    \label{fig:colour_dist}
\end{figure}

\Cref{fig:colour_dist} shows ($g - z$) colour distributions for four Fornax galaxies. 
We compare the original ACSFCS catalogue, the matched sources labelled as GCs after cross-referencing with the ACSFCS catalogue, and the sources our model, here shown for the CNN, identifies as GCs. 
For this comparison, we applied the average aperture corrections of $A_{g}$ = 0.237 mag and $A_z = 0.347$ as given in \cite{Jordan2009} to transform our m4\_g apertures to corrected magnitudes. 
We note that the choice of this aperture correction might slightly affect the colour distributions.
The colour distributions as recovered from the CNN model agree with the ACSFCS catalogue, especially for massive galaxies with a large number of GCs. 
For low-mass galaxies like FCC\,255 or FCC\,106 that only have a small number of GCs, the difference becomes larger due to low number statistics that is strongly affected by the classification of individual sources. 
Nonetheless, even in the extreme case of a galaxy like FCC\,106 with only 15 GCs, the recovered colour distribution is comparable to the ACSFCS result. Testing this for all Fornax galaxies shows that the mean colour from the ACSFCS catalogue is reproduced within 0.02 mag for systems with more than 100 GCs. For galaxies like FCC\,106 with less than 20 GCs, the mean colour is still within 0.12 mag from the ACSFCS value.
Likely, any results inferred from these distributions concerning, for instance, the assembly history would be similar for both distributions.

\section{Conclusions}
\label{sect:conclusions}
In this paper, we determine the performance of various machine learning techniques of varying complexity for the detection of GCs in archival HST data of the Fornax and Virgo galaxy clusters. We employ several methods on both image data and tabular data extracted from the image data using different evaluation strategies.
Our main results are:
\begin{itemize}
    \item Tree-based or hierarchical models like RFs and neural networks reach the best performances, outperforming more simple methods such as logistic regression, nearest neighbours classification, and (linear) support vector machines. 
    \item RFs on tabular data as well as CNNs on image data reach comparable performance levels, recovering $\sim 90-94$\% of GCs under test conditions, with only $\sim 6-8$\% of detected GCs being false positives.
    \item In the magnitude range 22 $<$ m4\_g $\leq$ 24.5 mag, the true positive rate even reaches 98 $-$ 99 \% with a false discovery rate $< 5$ \%.
    \item Although a lower false positive rate would be preferred, exploring false positives in more detail shows that at least some of those are actual GCs that were not labelled as such in the original catalogue.
    \item Our experiments show that models trained on data of one environment can be applied to data of other environments if similar conditions on galaxy types and distance are met.
    \item Explainable artificial intelligence methods provide additional ways of getting insight into the decision process of trained models as well as the training data. Using such methods shows that magnitudes, colours, and concentration indices are most important for identifying GCs, in agreement with traditional detection methods.
    \item As one exemplary science case, we compare GC colour distributions as derived from our best models to the distributions from the original catalogues. This shows that the distributions are recovered even at very low GC numbers ($<$ 20), indicating that science results derived from our detected sources will agree with the results from the original catalogues.
    \item The collected tabular and image data sets consisting of 18556 GCs and 84929 sources in total are made publicly available for future research.
\end{itemize}

This study gives an encouraging outlook of the potential machine learning might unfold when applied to future data of upcoming wide-field survey facilities such as Euclid or the Nancy Roman Space Telescope. 
The presented approach builds on the effort that has been put into collecting clean catalogues of a relatively small number of galaxies.
Given our results, we are confident that machine learning models trained by cross-referencing these catalogues will assist in creating novel catalogues of similar quality, but containing many more sources using less processed data. 
Combined with currently developed interpretable methods (and coupled with traditional approaches), such models promise to produce reliable and comprehensible results, which is essential if machine learning methods are to be applied to future data volumes too large for manual inspection.

\begin{acknowledgements}
We thank the referee Brad Whitmore for helpful comments and suggestions that helped to polish and improve this manuscript.
We thank Andreas Baumbach, Akos Kungl, and Guido De Marchi for valuable feedback on the manuscript. DD and KF acknowledge support through the European Space Agency fellowship programme. DD further thanks his colleagues at ESA’s Advanced Concepts Team for their ongoing support. Based on observations made with the NASA/ESA Hubble Space Telescope, and obtained from the Hubble Legacy Archive, which is a collaboration between the Space Telescope Science Institute (STScI/NASA), the Space Telescope European Coordinating Facility (ST-ECF/ESA) and the Canadian Astronomy Data Centre (CADC/NRC/CSA).
This research made use of Astropy,\footnote{\url{http://www.astropy.org}} a community-developed core Python package for Astronomy \citep{astropy2013, astropy2018}. This research made use of Photutils, an Astropy package for
detection and photometry of astronomical sources \citep{photutils}.
Furthermore, this research made use of the open source libraries Scikit-Learn \citep{scikit-learn}, PyTorch \citep{NEURIPS2019_9015}, NumPy \citep{harris2020array}, Matplotlib \citep{Hunter:2007}, pandas \citep{jeff_reback_2020_3715232,mckinney-proc-scipy-2010}, tqdm \citep{casper_da_costa_luis_2021_5517697} and CatBoost \citep{NEURIPS2018_14491b75}.
\end{acknowledgements}

\bibliographystyle{aa} 
\bibliography{References}

\begin{appendix}
\section{Used features}
\label{app:features}
We describe the features of the tabular data in the following. They were extracted using Photutil's segmentation\footnote{\url{https://photutils.readthedocs.io/en/stable/segmentation.html}} and aperture photometry routines after subtracting a median background from the HST ACS images of each galaxy. Not all features produced by these methods are used in the modelling, as for example the coordinates are not relevant.

\begin{itemize}
    \item m3\_g, m3\_z, m4\_g, m4\_z, m5\_g, and m5\_z are the aperture magnitudes in the two filters extracted using 3, 4, and 5 pixel radii, respectively.
    \item CI3\_g, CI3\_z, CI4\_g, CI4\_z, CI5\_g, and CI5\_z are the concentration indices corresponding to the magnitude difference between a 1 pixel aperture and 3, 4, or 5 pixels, respectively.
    \item colour refers to m4\_g - m4\_z.
    \item eccentricity, eccentricity\_z, orientation, orientation\_z are the eccentricity and orientation (position angle) of each source in the $g$ and $z$ bands, respectively.
    \item semimajor\_sigma, semiminor\_sigma, semimajor\_sigma\_z, and semiminor\_sigma\_z are measurements of the semimajor and semiminor axis lengths of each source.
    \item area and area\_z refer to the area of each source.
    \item min\_value, max\_value, min\_value\_z, and max\_value\_z are the minimum and maximum flux value of each source in the $g$ and $z$ bands, respectively.
    \item segment\_flux and segment\_flux\_z are the fluxes of each source as computed by Photutil's segmentation routines.
\end{itemize}

\section{Models}
\label{app:models}

For all models with learnable parameters we use L2 regularisation. 
For instance, for a parameter $b \in \mathbb{R}^{N}$, a term $\propto \sum_{j=1}^N\|b_j\|^2$ with $\| \cdot \|$ being the absolute value is added to the loss $\mathcal{L}$.

\subsection{Expanded notation}\label{sec:expanded_notation}

In the following, the used machine learning models are described as functions $f: \mathbb{R}^{N} \to \mathbb{R}$ that map from the feature space into the real space $x \mapsto f(x)$.
This outcome can be further squished by another function $g$, for example to obtain a probability estimate on which we can apply a decision rule to obtain the model prediction $y_\mathrm{m}(x)$ with $y_\mathrm{m}: \mathbb{R}^N \to \{0,1\}$,
\begin{equation}
    y_\mathrm{m}(x) = \begin{cases}
    1, & \text{if} \quad g\left(f(x)\right) > \theta \quad  \text{(source classified as a GC)}\,,\\
    0, & \text{else} \quad \text{(source not classified as a GC)}\,,
  \end{cases}
\end{equation}
where $\theta$ is a threshold value, for instance $\theta = 0.5\,$, and $g\left(f(x)\right) = p(x)$.
For the methods discussed here, $g$ is either a logistic function ($\sigma: \mathbb{R} \to [0,1]$)
\begin{equation}
    \sigma(x) = \frac{1}{1+\exp\left(-x\right)}   \,,  
\end{equation}
the sign function ($\mathrm{sign}: \mathbb{R} \to \{-1,1\}$)
\begin{equation}
    \mathrm{sign}(x) =  \begin{cases}
    1,  & \text{if\,} x \geq 0\,,\\
    -1, & \text{otherwise}\,,
    \end{cases}
\end{equation}
or the identity function.

\subsubsection{Logistic regression}

Given a feature vector $x \in \mathbb{R}^N$, logistic regression provides a probability estimate for its class affiliation, 
\begin{align}
    p(x\,\,|\,\,W\,,\,b) &= \sigma\big(f(x\,\,|\,\,W\,,\,b)\big) \,, \quad \text{with}\\
    f(x\,\,|\,\,W\,,\,b) &= W x + b \,,
\end{align}
where $W \in \mathbb{R}^{1\times N}$ and $b \in \mathbb{R}$ are learnable parameters.
To ease notation, we drop the dependence on $W$ and $b$ in the following.
An input $x$ is classified as a GC if $p(x) > 0.5\,$, which is identical to requiring $f(x) > 0\,$.
The parameters $W$ and $b$ are obtained by minimising a cross-entropy loss
\begin{equation}
    \mathcal{L} = \sum_{i} \yI \log\left(p\left(\xI\right)\right) + \left(1-\yI\right) \log\left(1-p\left(\xI\right)\right) \,. \label{eq:xentropy-loss}
\end{equation}

\subsubsection{Support vector machines}

Similar to logistic regression, a linear support vector machine classifies a source with feature vector $x \in \mathbb{R}^N$ as a GC if $\mathrm{sign}\big[f(x)\big] > 0$, with $W$ now given by 
\begin{equation}
    W = \sum_{i} w_i\, \zI \xI \,,
\end{equation}
where $w_i \geq 0 $ are learnable parameters and $\zI = 2\yI - 1$.
$W$ and $b$ are optimised using a hinge loss, 
\begin{equation}
    \mathcal{L} = \sum_{i} \mathrm{max}\left[0,\, 1 - \zI f\left(\xI\right) \right],
\end{equation}
which increases the margin between the found decision boundary and data points close to it.
To deal with non-linearly separable data, the input features $x$ can be mapped into a high-dimensional vector space $\varphi(x)$, resulting in
\begin{equation}
    f(x) = W \varphi(x) + b = \sum_{i} w_i\, \zI \varphi(\xI) \cdot \varphi(x) + b \,,
\end{equation}
which depends on the inner product of the mapped feature vectors.
The kernel trick states that these products can be reduced to inner products in the original feature space, weighted by some kernel $\kappa\left(\cdot, \cdot\right)$, allowing us to write our model as
\begin{equation}
    f(x) = \sum_{i} w_i\, \zI \kappa\left(\xI,\, x\right) + b \,.
\end{equation}
The linear support vector machine can be recovered by choosing $\kappa\left(\xI,\, x\right) = \xI \cdot x$, which only differs from logistic regression in the choice of the loss function.

\subsubsection{Neural networks}\label{app:neuralnet}

A neural network is a hierarchy of functional units
\begin{align}
    &f(x) = a^{(n)} \circ F(x) \,, \\
    &F(x) = f^{(n-1)} \circ \cdot \cdot \cdot \circ f^{(1)}(x) \,,
\end{align}
where $\circ$ is the concatenation of functions, $f^{(2)} \circ f^{(1)}(x) =f^{(2)}\left(f^{(1)}(x)\right)$.
The functional units $f^{(l)}$ are called the layers $l$ of the neural network and the individual elements of each layer are called neurons ($f^{(l)}_j$ is the output of neuron $j$ of layer $l$).
Each layer has a stereotypical structure, which in our case is given by
\begin{align}
    &f^{(l)}(x) = \delta \circ \phi \circ a^{(l)}(x) \,, \label{eq:f}\\
    &a^{(l)}(x) = W^{(l)} x + b^{(l)} \label{eq:a} \,, \\
    &\phi(x) = \mathrm{max}\left(0,x\right) \label{eq:phi} \,.
\end{align}
$\delta(x)$ is the dropout operator that randomly sets a fraction of elements of its input to zero during training to avoid overfitting \citep{srivastava2014dropout}.
The $W^{(l)} \in \mathbb{R}^{n^{(l)}\times n^{(l-1)}}$ are learnable weights that map from the $n^{(l-1)}$ neurons of layer $l-1$ to the $n^{(l)}$ neurons of layer $l$.
$b^{(l)} \in \mathbb{R}^{n^{(l)}}$ are learnable biases.
$\phi(x)$ is the so-called rectified linear unit (ReLU) activation function \citep{nair2010rectified} which is widely used in the literature.
In our case, a probabilistic estimate for the class of an input is obtained by applying a logistic activation function to the output, $p(x) = \sigma\left(f(x)\right)$.
Therefore, a source $x$ is classified as a GC if $p(x) > 0.5\,$.

The dropout operator $\delta$ is usually only applied during training, but can also be used during inference time \citep{gal2016dropout}.
In this case, evaluating $f(x)$ multiple times will result in a distribution of values for $p(x)$.
Thus, for each source, we can not only provide a class label via the mean of the sampled distribution, but through its width also a sensible estimate of the network's uncertainty (see \cref{app:dropout}).

The network parameters $W^{(l)}$ and $b^{(l)}$ are obtained $\forall l \in [0,n]$ by minimising a cross-entropy loss (\cref{eq:xentropy-loss}) using the error backpropagation algorithm.

\subsubsection{k nearest neighbours}

In the introduced notation, the decision function of kNN can be written as
\begin{equation}
    p(x) = \frac{1}{k} \sum_{i \in \mathcal{N}_k(x)} \yI,
\end{equation}
where $\mathcal{N}_k(x)$ is a set containing the indices of the k nearest neighbours of $x$ from the training data set.
A source is classified as a GC if $p(x) > 0.5$.

\subsubsection{Convolutional neural network}\label{app:CNN}

Here, we use the following architecture
\begin{align}
    &f(x) = a^{(5)} \circ F(x) \,,
\end{align}
with
\begin{align}
    &F(x) = f^{(4)} \circ \delta \circ c^{(3)} \circ \delta \circ P_\mathrm{max} \circ c^{(2)} \circ \delta \circ c^{(1)}(x) \,, \\
    &c^{(l)} = \phi\left(W^{(l)} * x + b^{(l)}\right) \,,
\end{align}
and $f^{(l)}$, $a^{(l)}$, $\phi$ and $\delta$ given by \cref{eq:f,eq:phi,eq:a}.
$c^{(l)}$ are convolutional layers, where $W^{(l)}$ is a set of learnable filters that are convolved with the input $x \in \mathbb{R}^{d\times M\times M}$.
$P_\mathrm{max}(x)$ is the max-pooling operator, which reduces the dimension of $x$ by replacing neighbouring pixels by a single pixel containing the maximum value.
For instance, a $2\times 2$ max-pooling operator returns $P_\mathrm{max}([0,1],[2,3]) = 3$.
A probabilistic estimate for the class of an input is obtained by applying a logistic activation function to the output, $p(x) = \sigma\left(f(x)\right)$.
Therefore, a source $x$ is classified as a GC if $p(x) > 0.5\,$.

\section{Feature rescaling}
\label{app:rescaling}

For the tabular data, features $k$ were rescaled by choosing $\alpha_k$ and $\beta_k$ as follows:
\begin{itemize}
    \item m3\_g, m3\_z, m4\_g, m4\_z, m5\_g, m5\_z, CI3\_g, CI3\_z, CI4\_g, CI4\_z, CI5\_g, CI5\_z and colour: $\alpha_k$ is given by the mean and $\beta_k$ by the standard deviation of all values of feature $k$ in the training data.
    \item orientation, orientation\_z, semiminor\_sigma, semiminor\_sigma\_z: $\alpha_k = 0$ and $\beta_k$ is given by the maximum value.
    \item area, area\_z, min\_value, min\_value\_z: $\alpha_k = 0$ and $\beta_k$ is given by the mean value.
    \item segment\_flux, segment\_flux\_z: $\alpha_k = 0$ and $\beta_k$ is given by the $90^\mathrm{th}$ percentile.
    \item max\_value, max\_value\_z: $\alpha_k = 0$ and $\beta_k$ is given by the $95^\mathrm{th}$ percentile.
    \item semimajor\_sigma, semimajor\_sigma\_z: $\alpha_k = 0$ and $\beta_k$ is given by the $99^\mathrm{th}$ percentile.
    \item eccentricity, eccentricity\_z: $\alpha_k = 0$ and $\beta_k = 1$.
\end{itemize}
The respective values were chosen by visually inspecting the distribution in the training split.

\section{Monte Carlo Dropout}
\label{app:dropout}

\begin{figure}
    \centering
    \includegraphics[width=\columnwidth]{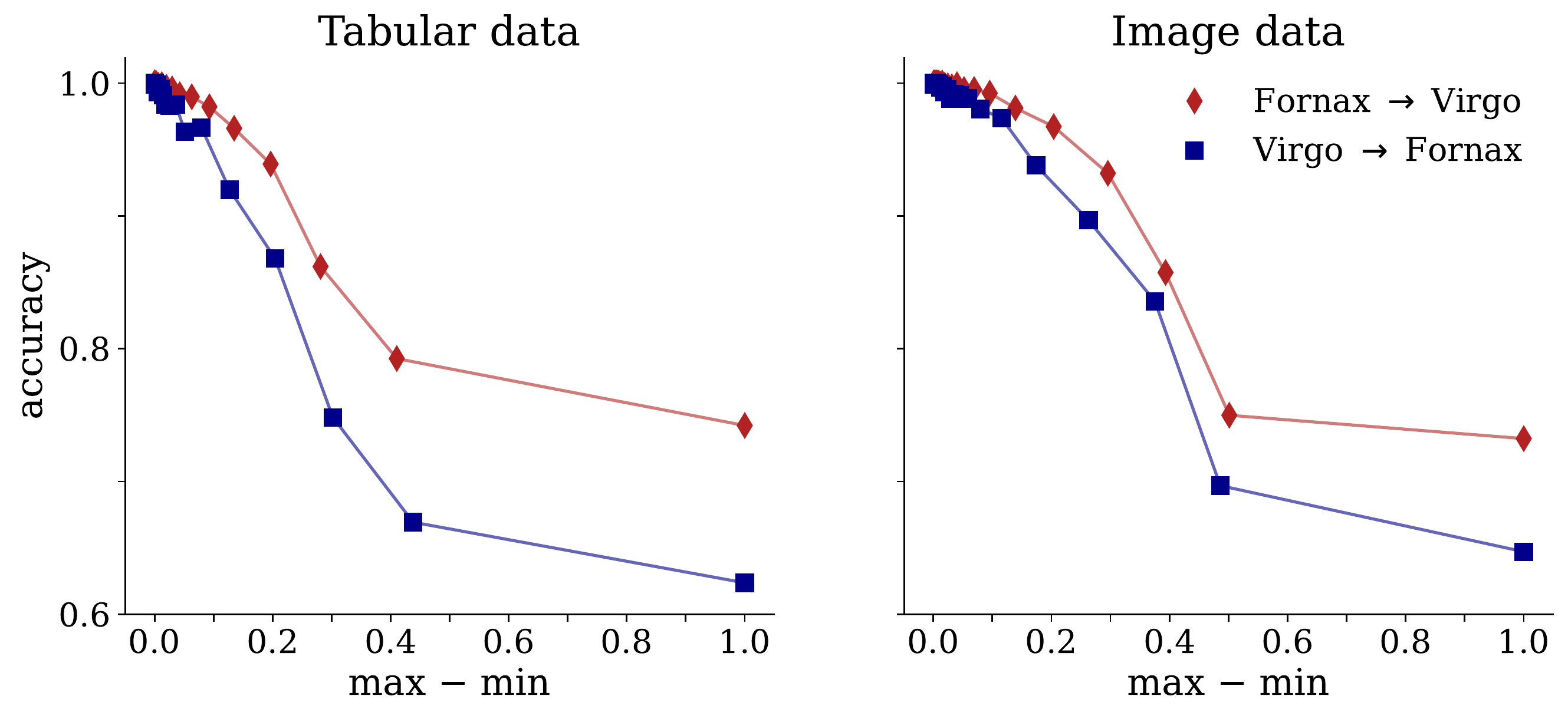}
    \caption{For higher uncertainties, model predictions become less accurate.
    Shown for a deep neural network trained on tabular data (left) and a CNN trained on image data (right).}
    \label{fig:CNN_uncertainty}
\end{figure}

By sampling repeatedly from a neural network with dropout, a distribution of model predictions is collected from which a measure of uncertainty can be derived. 
This is shown in \cref{fig:CNN_uncertainty} for a deep neural network trained on tabular data and a CNN trained on image data.
As a measure of uncertainty, we use the maximum span of the distribution, more specifically, the difference between the maximum and minimum score.
As required of a measure of uncertainty, for higher uncertainty ranges, the fraction of correctly classified sources (accuracy) drops.
In \cref{fig:CNN_uncertainty}, each data point corresponds to a bin of model predictions with uncertainties in a certain range, where each bin contains 4\% of all values.
In all simulations, we used a dropout probability of 5\%.

\section{Training only on Fornax data}
\label{app:tables}

\begin{table*}[!b]\renewcommand{\arraystretch}{1.2}
\caption{Training machine learning methods on tabular data of Fornax and evaluating it on Virgo data. }\vspace{-4mm}
\begin{center}
\begin{tabular}{c | c c c c | c c c c | c c}
\multicolumn{1}{c}{}&\multicolumn{4}{c}{averaged per galaxy}&\multicolumn{6}{c}{average over all sources}\\
\hline\hline
Method & TPR & FPR & FDR & AUC ROC & TPR & FPR & FDR & AUC ROC & \# TPs & \# FPs \\
\hline\hline
Logistic Regression & 0.73 & 0.15 & 0.58 & 0.88 & 0.83 & 0.16 & 0.44 & 0.91 & 10241 & 8079 \\
Support Vector Machine (linear) & 0.76 & 0.14 & 0.56 & -- & 0.84 & 0.14 & 0.41 & -- & 10405 & 7293  \\
Support Vector Machine (radial) & 0.83 & 0.05 & 0.33 & -- & 0.90 & 0.04 & 0.16 & -- & 11213 & 2165  \\
Nearest Neighbour & 0.85 & 0.04 & 0.28 & $-$ & 0.89 & 0.04 & 0.15 & $-$ & 11031 & 1882 \\
12 Nearest Neighbours & 0.87 & 0.03 & 0.22 & 0.98 & 0.91 & 0.03 & 0.10 & \best{0.99} & 11338 & 1300 \\
Decision Tree & 0.84 & 0.03 & 0.26 & 0.95 & 0.88 & 0.03 & 0.12 & 0.97 & 10889 & 1520 \\ 
Random Forest & 0.84 & \best{0.02} & \best{0.14} & 0.98 & 0.89 & \best{0.01} & \best{0.06} & \best{0.99} & 11086 & \best{698} \\
AdaBoost  & 0.84 & \best{0.02} & \best{0.14} & 0.98 & 0.89 & \best{0.01} & \best{0.06} & \best{0.99} & 11082 & 700\\
CatBoost & 0.85 & \best{0.02} & 0.17 & 0.98 & 0.90 & 0.02 & 0.08 & \best{0.99} & 11169 & 908 \\
Neural Network (29-1) & 0.73 & 0.17 & 0.60 & 0.87 & 0.82 & 0.19 & 0.49 & 0.89 & 10193 & 9746 \\
Neural Network (29-100-100-1) & \best{0.88} & \best{0.02} & 0.17 & \best{0.99} & \best{0.92} & 0.02 & 0.08 & \best{0.99} & \best{11443} & 937  \\
\hline
\end{tabular}
\label{tab:featureLearningFornax2Virgo}
\end{center}
\tablefoot{The Virgo data contains in total 63162 sources with 12395 catalogued GCs. Results are given for a decision threshold of $0.5$. Similar to Table \ref{tab:featureLearningVirgo2Fornax}.}
\end{table*}
\begin{table*}[!b]\renewcommand{\arraystretch}{1.2}
\caption{Training machine learning methods on image data of Fornax and evaluating it on Virgo image data.}\vspace{-4mm}
\begin{center}
\begin{tabular}{c | c c c c | c c c c | c c}
\multicolumn{1}{c}{}&\multicolumn{4}{c}{averaged per galaxy}&\multicolumn{6}{c}{average over all sources}\\
\hline\hline
Method & TPR & FPR & FDR & AUC ROC & TPR & FPR & FDR & AUC ROC & \# TPs & \# FPs \\
\hline\hline
Nearest Neighbour & 0.81 & 0.04 & 0.31 & $-$ & 0.86 & 0.04 & 0.15 & $-$ & 10613 & 1926  \\
12 Nearest Neighbours & 0.80 & 0.04 & 0.31 & 0.96 & 0.86 & 0.04 & 0.15 & 0.97 & 10676 & 1815 \\
Random Forest & 0.77 & 0.04 & 0.32 & 0.95 & 0.85 & 0.04 & 0.15 & 0.97 & 10468 & 1823  \\
\hline
Convolutional Neural Network (CNN) & \best{0.90} & 0.03 & 0.20 & \best{0.99} & \best{0.93} & \best{0.02} & 0.09 & \best{0.99} & \best{11560} & 1194  \\
CNN + Nearest Neighbour & 0.84 & 0.03 & 0.20 & $-$ & 0.88 & \best{0.02} & 0.09 & $-$ & 10920 & 1132  \\
CNN + 12 Nearest Neighbours & 0.86 & \best{0.02} & \best{0.15} & 0.98 & 0.91 & \best{0.02} & \best{0.06} & \best{0.99} & 11239 & \best{770} \\
\hline\hline
\end{tabular}
\label{tab:imageLearningFornax2Virgo}
\end{center}
\tablefoot{The Virgo data contains in total 62581 sources (we excluded sources from VCC538 as it contained a lot of false positives from a background galaxy)  with 12385 catalogued GCs. Results are given for a decision threshold of $0.5$. Similar to \cref{tab:imageLearningVirgo2Fornax}.}
\end{table*}
Tables \ref{tab:featureLearningFornax2Virgo} and \ref{tab:imageLearningFornax2Virgo} show the model performances when training on the smaller data set of Fornax and evaluating it on Virgo. The performances are comparable to those shown in Tables \ref{tab:featureLearningVirgo2Fornax} and \ref{tab:imageLearningVirgo2Fornax} when training on Virgo and evaluating on Fornax.

\section{ROC curves}
\label{app:roc}

\begin{figure}[h]
    \centering
    \includegraphics[width=\columnwidth]{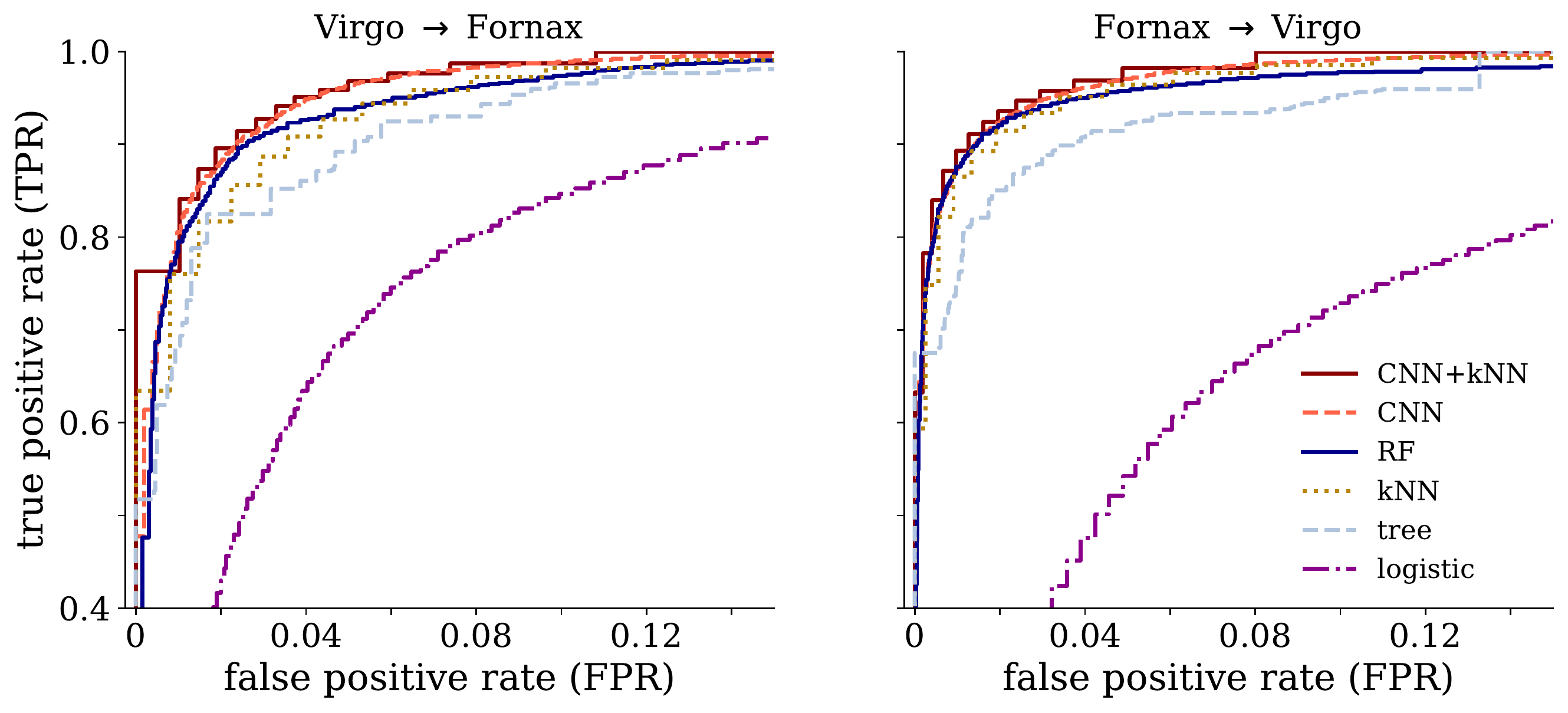}
    \caption{ROC curves for several models and (left) training on Virgo and testing on Fornax data and (right) vice versa. In both cases, CNN+kNN as well as CNN show the best characteristics, closely followed by RFs. To increase readability, we show a zoom-in here with false positives rates ranging from $0$ to $0.14$ and true positives rates from $0.4$ to $1$.}
    \label{fig:roc}
\end{figure}

ROC curves for several models and the evaluation scenario where training and test data is split between galaxy clusters are shown in \cref{fig:roc}. Apart from CNN+kNN and CNN, all models were trained on tabular data. For kNN, we used $k=12$ here.

\section{Dependence on magnitudes}\label{app:magnitude}

\begin{figure}
    \centering
    \includegraphics[width=0.925\columnwidth]{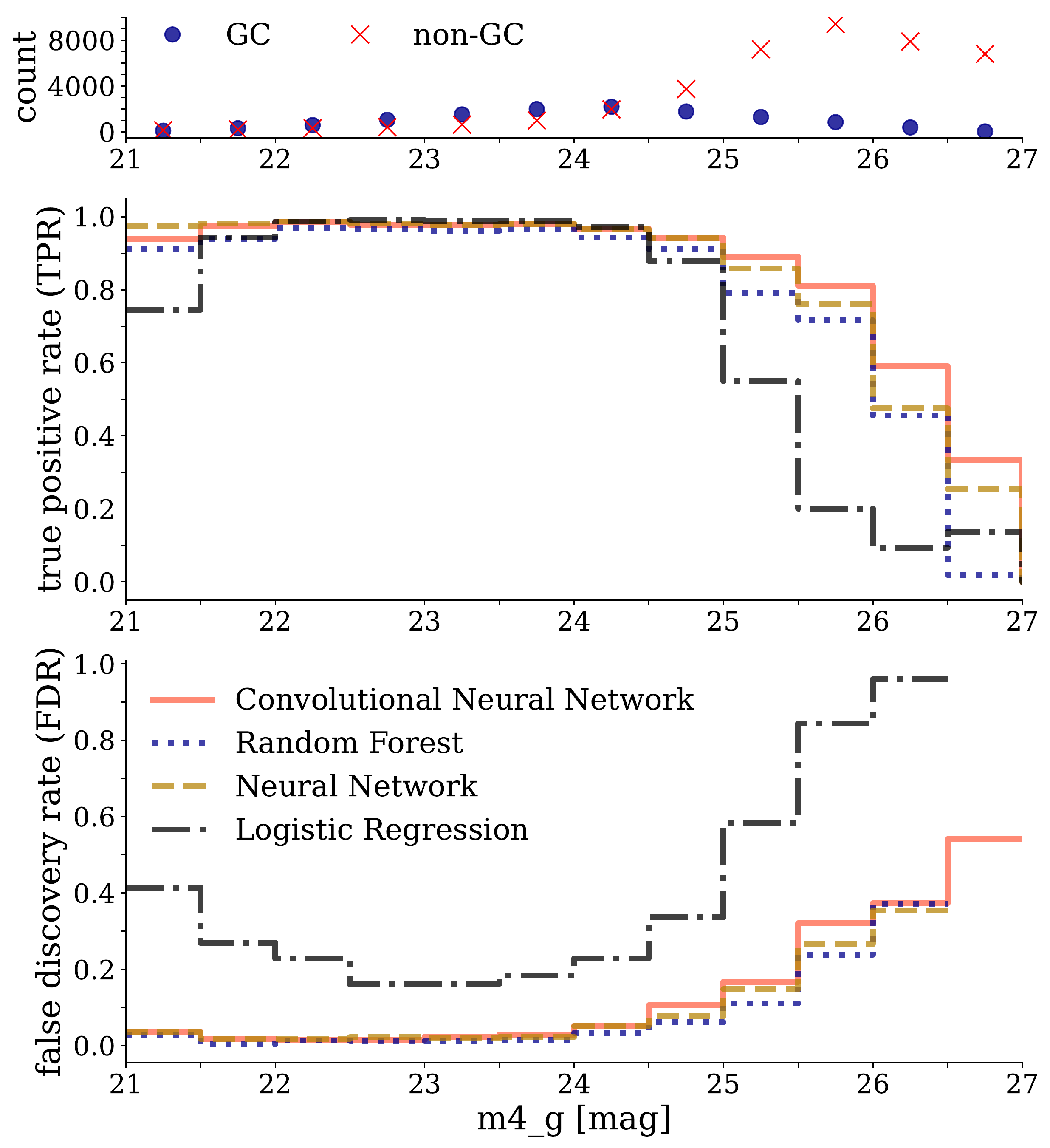}
    \caption{Model performance for sources of different magnitude ranges, separated into bins of size $0.5$ mag, for the case of training on data of the Fornax cluster and evaluating on data of Virgo.}
    \label{fig:magnitude_dependence_appendix}
\end{figure}

True positive rates as well as false discovery rates for sources of different magnitudes are shown in \cref{fig:magnitude_dependence_appendix} for the case where models are trained on Fornax data and evaluated on Virgo data.
In this case, the performance drop for fainter sources is more pronounced, since the training data contained mostly brighter sources.

\section{Evaluating LIME for multiple GCs}\label{app:limeExp}

\begin{figure}[h]
    \centering
    \includegraphics[width=0.75\columnwidth]{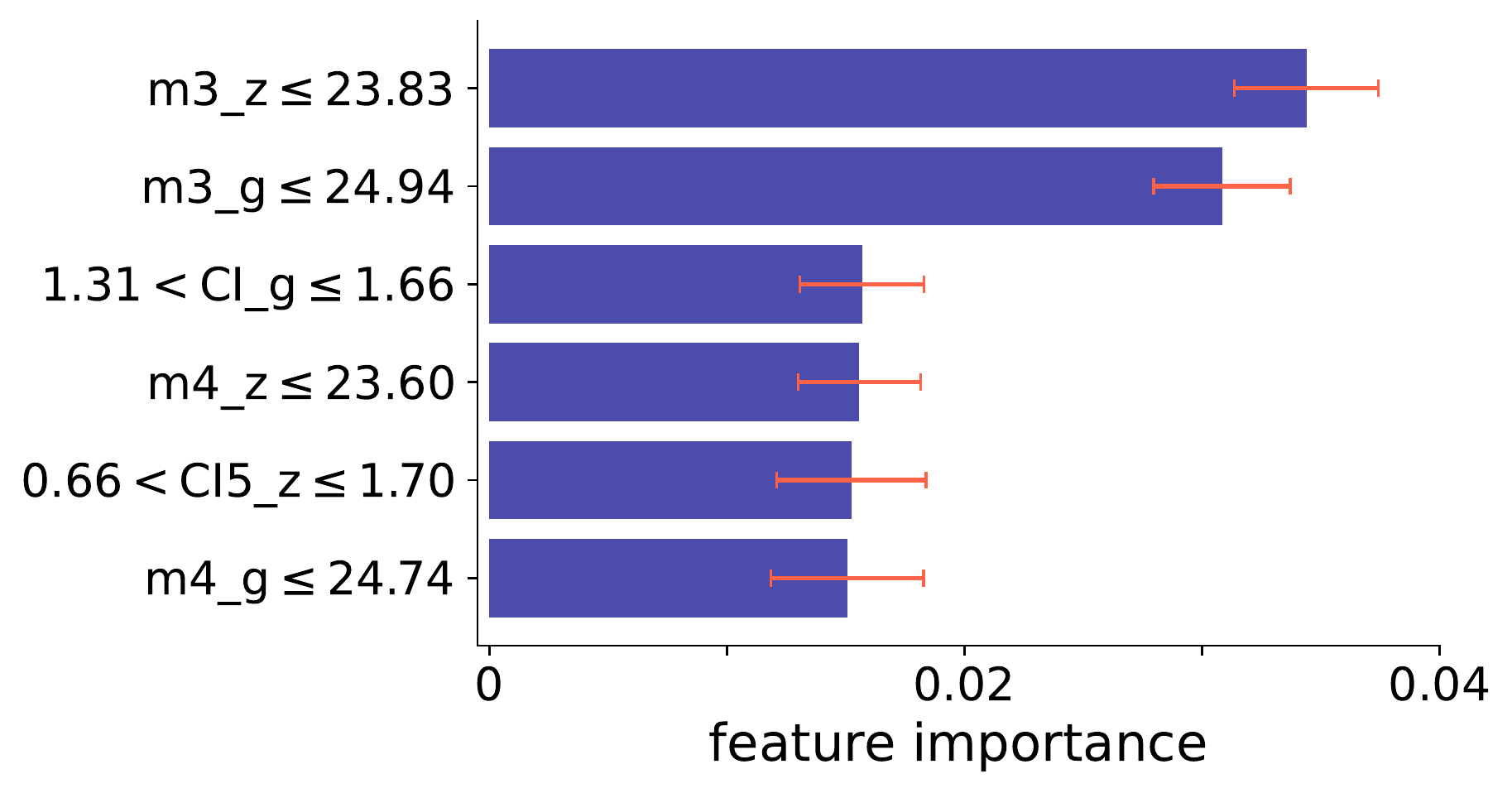}
    \caption{Most important features for identifying GCs according to LIME. In blue, mean values for the feature importance are given, calculated from explanations generated for the 345 correctly classified GCs in FCC219. Red error bars denote standard deviation.}
    \label{fig:fcc219LIME}
\end{figure}

We used LIME to generate explanations for all GCs correctly classified with the RF method in the Fornax galaxy FCC219, which amounts to 345 GCs in total.
In \cref{fig:fcc219LIME}, the average of all explanations for these GCs is shown, demonstrating that (i) according to LIME, similar criteria are applied to identify different GCs and (ii) the most important features for detecting GCs extracted by the model are consistent with those traditionally used to find GCs.

\end{appendix}

\end{document}